\documentclass[aps,pra,superscriptaddress,twocolumn]{revtex4-2}

\usepackage{graphicx}
\usepackage{multirow}
\usepackage{amsmath,amssymb,amsfonts}
\usepackage{amsthm}
\usepackage{mathrsfs}
\usepackage{xcolor}
\usepackage{textcomp}
\usepackage{booktabs}
\usepackage{algorithm}
\usepackage{algorithmicx}
\usepackage{algpseudocode}
\usepackage{listings}
\usepackage{dirtytalk}
\usepackage{xr}
\usepackage{braket}
\usepackage[hidelinks]{hyperref}
\usepackage[separate-uncertainty = true, multi-part-units=single]{siunitx}
\usepackage{placeins}

\usepackage{circuitikz} 
\usetikzlibrary{positioning, shapes.geometric, calc, decorations.pathmorphing, backgrounds, fit}

\newcommand{\cmark}{\textcolor{green!60!black}{\checkmark}}
\newcommand{\xmark}{\textcolor{red!70!black}{$\times$}}
\newcommand{\pmark}{\textcolor{orange!80!black}{$\circ$}}

\newif\ifequalcontribauthor
\DeclareRobustCommand{\equalcontrib}{\equalcontribauthortrue}

\makeatletter
\def\doauthor#1#2#3{%
  \equalcontribauthorfalse
  \ignorespaces#1\unskip\@listcomma
  \begingroup
    #3%
  \endgroup
  {\def\@authornoteseparator{\comma@space}}%
  {\let\@authornoteseparator\@empty}%
  \ifequalcontribauthor
    \@authornoteseparator
    \frontmatter@footnote{These authors contributed equally.}%
    \def\@authornoteseparator{\comma@space}%
  \fi
  \@if@empty{#2}{}{%
    \@authornoteseparator
    \frontmatter@footnote{#2}%
  }%
  \space\@listand
}
\makeatother

\tikzset{
    thick,
    dot/.style={circle, draw, minimum size=1.05cm, inner sep=0pt, fill=white, font=\large},
    resistor/.style={rectangle, draw, minimum width=0.38cm, minimum height=0.7cm, inner sep=0pt, fill=white}
}

\begin{document}

\title{QArray+: A physics-informed GPU-accelerated simulator for quantum dot arrays}

\author{Pranav Vaidhyanathan\equalcontrib}
\affiliation{NVIDIA, Santa Clara, California 95051, USA}
\affiliation{Department of Engineering Science, University of Oxford, Oxford OX1 3PJ, United Kingdom}

\author{Barnaby van Straaten\equalcontrib}
\email{b.l.vanstraaten@tudelft.nl}
\affiliation{QuTech and Kavli Institute of Nanoscience, Delft University of Technology, P.O. Box 5046, 2600 GA Delft, The Netherlands}

\author{Alice Petrillo}
\affiliation{QuTech and Kavli Institute of Nanoscience, Delft University of Technology, P.O. Box 5046, 2600 GA Delft, The Netherlands}

\author{Rahul Marchand}
\affiliation{Department of Engineering Science, University of Oxford, Oxford OX1 3PJ, United Kingdom}

\author{Edwin De Nicolo}
\affiliation{Department of Engineering Science, University of Oxford, Oxford OX1 3PJ, United Kingdom}

\author{Menno Veldhorst}
\affiliation{QuTech and Kavli Institute of Nanoscience, Delft University of Technology, P.O. Box 5046, 2600 GA Delft, The Netherlands}

\author{Brucek Khailany}
\affiliation{NVIDIA, Santa Clara, California 95051, USA}

\author{Taylor L. Patti}
\affiliation{NVIDIA, Santa Clara, California 95051, USA}

\author{Natalia Ares}
\email{natalia.ares@eng.ox.ac.uk} 
\affiliation{Department of Engineering Science, University of Oxford, Oxford OX1 3PJ, United Kingdom}

\date{\today}

\begin{abstract} 
Semiconductor quantum-dot arrays are a compelling platform for scalable quantum technologies, yet their practical operation is hindered by the complexity of tuning large-scale devices. Existing automation tools rely on simplified physical models---such as constant-capacitance approximations and equilibrium Hubbard models---which assume instantaneous relaxation to a steady state. These frameworks fail in experimentally critical regimes where measurement rates exceed tunneling dynamics, necessitating more sophisticated non-equilibrium control strategies. To bridge this gap, we introduce \texttt{QArray+}, an extension of the \texttt{QArray} framework that incorporates gate-dependent tunnel coupling and a quantum open-system description of dissipative processes. This approach enables the unified simulation of coherent interdot charge-state hybridization and the non-equilibrium latching dynamics essential for training robust machine-learning models for automated device operation. Implemented in JAX with GPU acceleration, \texttt{QArray+} scales across GPUs and multi-node systems. 
For example, a charge stability diagram for a $100 \times 100$ grid of  gate-voltages over $64$ dots can be computed in $\sim0.17\,\mathrm{s}$ on multiple GPUs.
Since interdot interactions are short-ranged and the corresponding tuning corrections are local, simulations at these scales capture the physics relevant to even larger devices. These capabilities support high-throughput dataset generation for automated device tuning.
\end{abstract}
\maketitle

\section{Introduction}

Semiconductor quantum dots (QDs) are a compelling platform for scalable quantum technologies, offering high-fidelity qubit operations and compatibility with modern semiconductor fabrication \cite{George_2024, steinacker2024300mmfoundrysilicon, nickl2025eightqubitoperation300mm, chittockwood2025radiofrequencycascadereadoutcoupled}. The realization of high-fidelity quantum circuits based on large QD arrays requires their tuning, control, and operation to be automated. To this end, machine-learning-based strategies for charge tuning have shown strong promise, but their development depends on realistic simulation environments and large, labeled datasets \cite{lidiak2025virtualgatesenableddigital, marchand2025endtoendanalysischargestability, Rao_2025, carlsson2025automatedallrftuningspin, Zwolak_2020, buterakos2025qdflowpythonpackagephysics, Moon_2020, Kalantre_2019, kovach2025bootstrappingautonomoustestinginitialization, Zwolak_2024}. Since acquiring such datasets experimentally is impractical, fast physics-based simulators have become an appealing method for creating automated tuning pipelines \cite{van_Straaten_2024, murphy2025rfsquadradiofrequencysimulatorquantum, Krzywda_2025, Gualtieri_2025, schorling2025meta, schuff2026fully, Moon_2020, carlsson2025automatedallrftuningspin}.

Existing tools, typically based on constant-capacitance models or simplified Hubbard models, have proven effective for generating charge-stability diagrams (CSDs) used in the training of ML models \cite{Rao_2025, carlsson2025automatedallrftuningspin, marchand2025endtoendanalysischargestability}. These simulators generate charge-stability diagrams by determining the system's steady-state charge configuration, usually assumed to be the one that minimizes the state's electrostatic energy at each point in gate-voltage space. However, this implicitly assumes that the dots' charge configuration can relax to that steady state on timescales much faster than the measurement timescale. This assumption breaks down whenever charge relaxation is slow on the measurement timescale. In particular, the coupling of a dot to its reservoir is exponentially suppressed with distance, so that relaxation times can become extremely long, leading to slow charge dynamics and hysteresis~\cite{borsoi2022sharedcontrol16semiconductor, vjohn2025, George_2024, De_Smet_2025, Yang_2014}. Overcoming these limitations requires simulators capable of capturing a broader range of physical phenomena while remaining computationally efficient. Specifically, advanced frameworks must be able to simulate non-equilibrium latching, coherent interdot tunnel coupling for charge hybridization, and the simultaneous presence of both via an open-quantum-system approach. Furthermore, realistic modeling demands accommodating for effects such as non-linear electrostatics through voltage-dependent capacitances, supporting isolated modes where dots are entirely decoupled from reservoirs, and utilizing GPU acceleration to keep these complex, large-scale calculations tractable. 

\begin{table*}
\centering
\begin{tabular}{lccccccc}
\toprule
 & \multicolumn{7}{c}{\textbf{Framework}} \\
\cmidrule(lr){2-8}
\textbf{Feature}
 & \textbf{QDSim}
 & \textbf{QArray}
 & \textbf{QDFlow}
 & \textbf{QDarts}
 & \textbf{SimCATS}
 & \textbf{RF-Squad}
 & \textbf{QArray+} \\
\midrule
Simulate tunnel coupling                & \xmark & \xmark & \xmark & \cmark & \pmark & \cmark & \cmark \\ 
Simulate latching                       & \xmark & \pmark & \pmark & \xmark & \xmark & \xmark & \cmark \\
Simulate latching and tunnel coupling     & \xmark & \xmark & \xmark & \xmark & \xmark & \xmark & \cmark \\
Simulate voltage-dependent capacitances & \xmark & \xmark & \cmark & \cmark & \xmark & \cmark & \cmark \\
Simulate isolated mode                  & \xmark & \cmark & \xmark & \xmark & \xmark & \xmark & \cmark \\
Simulate RF-reflectometry readout       & \xmark & \xmark & \xmark & \xmark & \xmark & \cmark & \xmark \\
Utilize GPU acceleration               & \xmark & \cmark & \xmark & \xmark & \xmark & \cmark & \cmark \\
Utilize multi-GPU acceleration               & \xmark & \xmark & \xmark & \xmark & \xmark & \xmark & \cmark \\
\bottomrule
\end{tabular}

\caption{Comparison of quantum dot simulation frameworks
(QDSim~\cite{Gualtieri_2025},
QArray~\cite{van_Straaten_2024},
QDFlow~\cite{buterakos2025qdflowpythonpackagephysics},
QDarts~\cite{Krzywda_2025},
SimCATS~\cite{Hader_2024},
RF-Squad~\cite{murphy2025rfsquadradiofrequencysimulatorquantum}).
Here, ``tunnel coupling'' denotes coherent interdot coupling that hybridizes charge states and broadens CSD transition lines, as opposed to microscopic lead-barrier transport, and ``latching'' denotes metastable non-equilibrium charge retention on scan timescales. The symbol $\circ$ indicates a phenomenological implementation that reproduces the qualitative feature without the underlying physics: for latching, the simplified relaxation model originally introduced in QArray, which neglects the gate-voltage dependence of the relaxation rate and the possibility of relaxation through intermediate charge states; for tunnel coupling, the geometric rounding and interpolation of interdot transitions in SimCATS, which yields broadened transition lines and fractional occupations without a coherent Hamiltonian. See Appendix~\ref{app:comparison} for a comparison of \texttt{QArray} with the old latching implementation and \texttt{QArray+}.}
\label{tab:comparison}
\end{table*}

\begin{figure}
	\centering
	\includegraphics{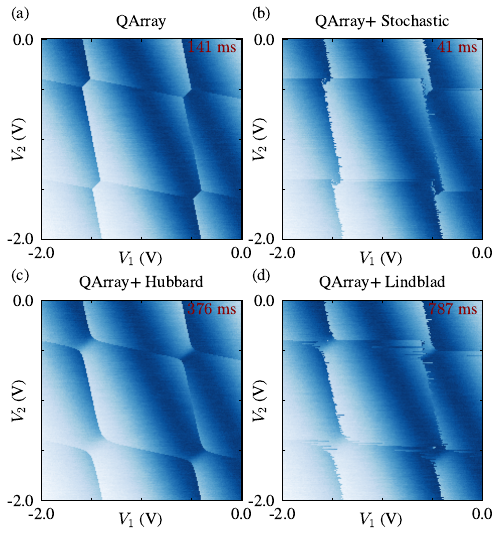}
    \caption{\textbf{Double dot charge stability diagram computed using four different models.} a) The steady state charge stability diagram computed using \texttt{QArray}. b-d) The latched charge stability diagram computed using the \texttt{QArray+} stochastic, spinless Hubbard and open quantum system models, respectively. The stochastic CSD shows latching but not hybridization of charge states. The spinless Hubbard model shows just hybridization of charge states. The open quantum system model shows both effects. The CPU compute time for each $200 \times 200$ CSD is indicated in the top-right corner.}
    \label{fig:models}
\end{figure}

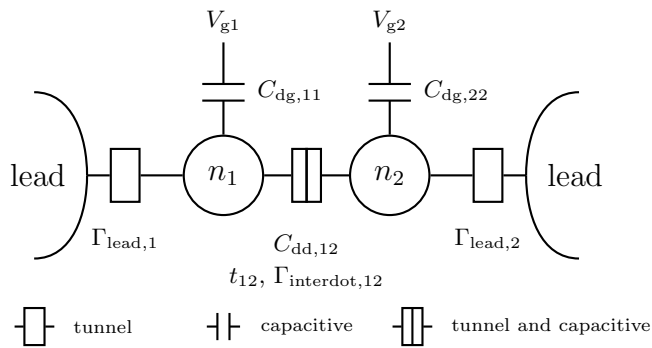
\begin{figure}
\centering
\begin{tikzpicture}[thick]

\node[dot] (N1) at (0,0) {$n_1$};
\node[dot] (N2) at (2.2,0) {$n_2$};

\node[resistor] (RL) at (-1.3,0) {};
\node[resistor] (Rm) at ($(N1)!0.5!(N2)$) {};
\draw[line width=1pt] (Rm.north) -- (Rm.south);
\node[resistor] (RR) at (3.5,0) {};

\draw (RL) -- (N1);
\draw (N1) -- (Rm);
\draw (Rm) -- (N2);
\draw (N2) -- (RR);

\draw (RL.west) -- ++(-0.3,0) coordinate (S_conn);
\draw[thick] (S_conn) to[out=90, in=0] ++(-0.7,1.1);
\draw[thick] (S_conn) to[out=-90,in=0] ++(-0.7,-1.1);
\node at ($(S_conn)+(-0.65,0)$) {\large lead};

\draw (RR.east) -- ++(0.3,0) coordinate (D_conn);
\draw[thick] (D_conn) to[out=90, in=180] ++(0.7,1.1);
\draw[thick] (D_conn) to[out=-90,in=180] ++(0.7,-1.1);
\node at ($(D_conn)+(0.65,0)$) {\large lead};

\node[below=0.22cm of RL] {$\Gamma_{\text{lead},1}$};
\node[below=0.22cm of Rm] {$$\begin{tabular}{c}
$C_{\mathrm{dd,12}}$ \\
$t_{12},\,\Gamma_{\text{interdot},12}$ \\
\end{tabular}$$};
\node[below=0.22cm of RR] {$\Gamma_{\text{lead},2}$};

\coordinate (Cg1_start) at ($(N1.north)+(0,0.45)$);
\draw (N1.north) -- (Cg1_start);
\draw[thick] ($(Cg1_start) + (-0.28,0)$) -- ($(Cg1_start) + (0.28,0)$);
\draw[thick] ($(Cg1_start) + (-0.28,0.22)$) -- ($(Cg1_start) + (0.28,0.22)$);
\draw ($(Cg1_start)+(0,0.22)$) -- ++(0,0.55) node[above] {$V_{\mathrm{g1}}$};
\node[right] at ($(Cg1_start)+(0.33,0.11)$) {$C_{\mathrm{dg,11}}$};

\coordinate (Cg2_start) at ($(N2.north)+(0,0.45)$);
\draw (N2.north) -- (Cg2_start);
\draw[thick] ($(Cg2_start) + (-0.28,0)$) -- ($(Cg2_start) + (0.28,0)$);
\draw[thick] ($(Cg2_start) + (-0.28,0.22)$) -- ($(Cg2_start) + (0.28,0.22)$);
\draw ($(Cg2_start)+(0,0.22)$) -- ++(0,0.55) node[above] {$V_{\mathrm{g2}}$};
\node[right] at ($(Cg2_start)+(0.33,0.11)$) {$C_{\mathrm{dg,22}}$};

\begin{scope}[shift={(0,-2.0)}]

    \node[resistor, scale=0.7] (LR) at (-2.5,0) {};
    \draw (LR.west)--++(-0.12,0);
    \draw (LR.east)--++(0.12,0);
    \node[right=0.25cm of LR] {\scriptsize tunnel};

    \node at (0,0) (LC) {};
    \draw ($(LC)+(-0.22,0)$) -- ($(LC)+(-0.08,0)$);
    \draw ($(LC)+(0.22,0)$) -- ($(LC)+(0.08,0)$);
    \draw ($(LC)+(-0.08,-0.18)$)--($(LC)+(-0.08,0.18)$);
    \draw ($(LC)+( 0.08,-0.18)$)--($(LC)+( 0.08,0.18)$);
    \node[right=0.25cm of LC] {\scriptsize capacitive};

    \node[resistor, scale=0.7] (LB) at (2.5,0) {};
    \draw (LB.north)--(LB.south);
    \draw (LB.west)--++(-0.12,0);
    \draw (LB.east)--++(0.12,0);
    \node[right=0.25cm of LB] {\scriptsize tunnel and capacitive};
\end{scope}

\end{tikzpicture}
\caption{\textbf{Schematic of a quantum dot device with tunnel and capacitive couplings.} Each quantum dot carries charge $n_i$ and is tunnel-coupled to the source and drain leads with rates $\Gamma_{\mathrm{lead},i}$. The dots are mutually coupled through the inter-dot capacitance $C_{dd,12} = C_{dd,21}$, the coherent inter-dot tunnel coupling $t_{12}$, and the incoherent inter-dot transition rate $\Gamma_{\mathrm{interdot},12}$. Gate electrodes at voltages $V_{g i}$ are capacitively coupled to the dots; for clarity, any cross-capacitances are omitted. The legend indicates the symbols used to represent tunnel, capacitive, and combined tunnel-capacitive couplings. Figure adapted from Fig.~1 of \cite{RevModPhys.75.1}.}
\label{fig: schematic}
\end{figure}

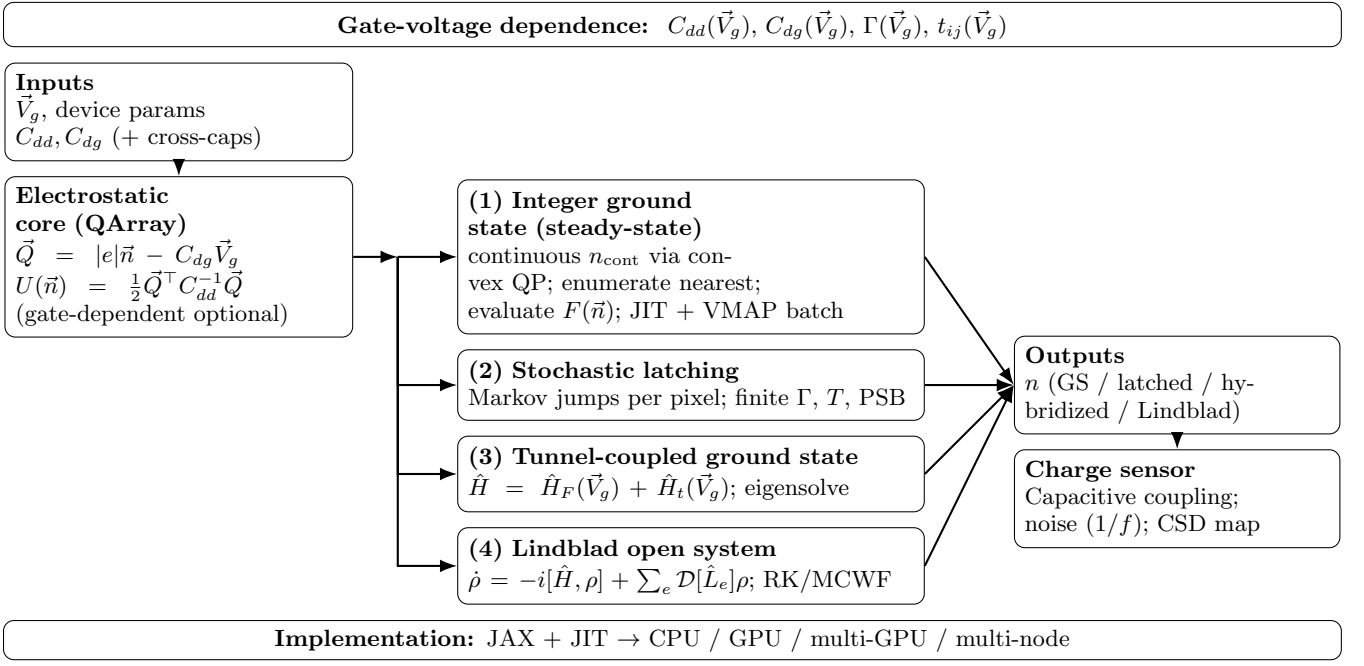
\begin{figure*}[t]
\centering
\resizebox{\textwidth}{!}{%
\begin{tikzpicture}[
  transform shape,
  font=\small,
  arr/.style={-Latex, thick},
  box/.style={draw, rounded corners, align=left, inner sep=4pt, text width=4.4cm},
  head/.style={draw, rounded corners, align=left, inner sep=4pt, text width=6.0cm},
  band/.style={draw, rounded corners, align=center, inner sep=3pt},
  node distance=2mm and 10mm %
]

\node[box] (inp) {
  \textbf{Inputs}\\
  $\vec V_g$, device params\\
  $C_{dd}, C_{dg}$ (+ cross-caps)
};

\node[box, below=of inp] (core) {
  \textbf{Electrostatic core (QArray)}\\
  $\vec Q=|e|\vec n - C_{dg}\vec V_g$\\
  $U(\vec n)=\tfrac12 \vec Q^\top C_{dd}^{-1}\vec Q$\\
  (gate-dependent optional)
};

\draw[arr] (inp) -- (core);

\node[head, right=14mm of core] (steady) {
  \textbf{(1) Integer ground state (steady-state)}\\
  continuous $n_{\mathrm{cont}}$ via convex QP; enumerate nearest;\\
  evaluate $F(\vec n)$; JIT + VMAP batch
};

\node[head, below=of steady] (latch) {
  \textbf{(2) Stochastic latching}\\
  Markov jumps per pixel; finite $\Gamma$, $T$, PSB
};

\node[head, below=of latch] (tunnel) {
  \textbf{(3) Tunnel-coupled ground state}\\
  $\hat H=\hat H_F(\vec V_g)+\hat H_t(\vec V_g)$; eigensolve
};

\node[head, below=of tunnel] (lind) {
  \textbf{(4) Lindblad open system}\\
  $\dot\rho=-i[\hat H,\rho]+\sum_e\mathcal{D}[\hat L_e]\rho$; RK/MCWF
};

\coordinate (split) at ($(core.east)+(6mm,0)$);
\draw[arr] (core.east) -- (split);
\draw[arr] (split) |- (steady.west);
\draw[arr] (split) |- (latch.west);
\draw[arr] (split) |- (tunnel.west);
\draw[arr] (split) |- (lind.west);

\node[box, right=12mm of latch, text width=4.1cm] (out) {
  \textbf{Outputs}\\
  $n$ (GS / latched / hybridized / Lindblad)
};

\node[box, below=2mm of out, text width=4.1cm] (sensor) {
  \textbf{Charge sensor}\\
  Capacitive coupling; noise ($1/f$); CSD map
};

\draw[arr] (steady.east) -- (out.west);
\draw[arr] (latch.east) -- (out.west);
\draw[arr] (tunnel.east) -- (out.west);
\draw[arr] (lind.east) -- (out.west);
\draw[arr] (out) -- (sensor);

\node[fit=(inp)(core)(steady)(latch)(tunnel)(lind)(out)(sensor),
      draw=none, inner sep=0pt] (full) {};

\path let
  \p1 = (full.east),
  \p2 = (full.west),
  \n1 = {\x1-\x2-2mm}
in
  node[band, above=2mm of full.north, text width=\n1] (gvdep) {%
    \textbf{Gate-voltage dependence:}\;
    $C_{dd}(\vec V_g),\, C_{dg}(\vec V_g),\, \Gamma(\vec V_g),\, t_{ij}(\vec V_g)$
  }
  node[band, below=2mm of full.south, text width=\n1] (impl) {%
    \textbf{Implementation:} JAX + JIT $\rightarrow$ CPU / GPU / multi-GPU / multi-node
  };

\end{tikzpicture}
}
\caption{Overview of QArray+ relative to QArray: shared electrostatic core with four interchangeable simulation heads (stochastic latching, tunnel-coupled ground state, Lindblad open-system evolution, and integer ground-state via continuous relaxation + candidate enumeration + energy selection), plus shared gate-voltage dependence, readout/noise, and JAX GPU acceleration.}
\label{fig: structure}
\end{figure*}

We introduce \texttt{QArray+}, a successor to \texttt{QArray} \cite{van_Straaten_2024}. By incorporating both interdot tunnel coupling and finite reservoir tunneling rates, \texttt{QArray+} extends the original capacitance-based framework to capture a broader range of experimentally observable phenomena, such as coherent interdot hybridization and non-equilibrium charge dynamics. This physics-informed simulator is explicitly designed for acceleration and scalability. Because it is implemented in JAX with just-in-time compilation and GPU execution, it is easily executable in a single GPU as well as multi-GPU and multi-node environments. Furthermore, its electrostatic core uses a polynomial reformulation that reduces the per-point compute cost of generating charge-stability diagrams, ultimately enabling dataset-scale synthesis.  The package provides three complementary dynamical models for computing CSDs, each with differing trade-offs between computational complexity and the captured physical phenomena:
\begin{itemize}
    \item[-] \textbf{Stochastic Capacitance Model:} This approach treats charge dynamics as classical Markov jump processes between discrete integer occupation states. Within this framework, the model effectively captures non-equilibrium latching phenomena, essential for simulating regimes where transition rates are slow relative to readout timescales. 
    
    \item[-] \textbf{Spinless Hubbard Model:} To account for quantum mechanical effects in the steady state, this model implements a spinless Hubbard Hamiltonian. This allows for the computation of CSDs that exhibit coherent interdot hybridization, though it assumes a steady-state equilibrium that precludes the observation of non-equilibrium latching, mirroring the capabilities introduced in Ref. \cite{murphy2025rfsquadradiofrequencysimulatorquantum} (see \autoref{tab:comparison}).
        
    \item[-] \textbf{Open-Quantum-System Model:} To provide a unified simulation of both coherent and incoherent processes, this model describes the array with a Lindblad master equation combining phonon and lead dissipators, which is evolved as a stochastic quantum-jump (Monte Carlo wavefunction) trajectory in the instantaneous eigenbasis. By bridging the gap between the previous two approaches, it enables the simultaneous treatment of coherent hybridization and the incoherent tunneling processes that lead to latching, offering a complete description of the transition between these regimes.
\end{itemize}
\autoref{tab:comparison} summarizes the capabilities of existing simulators along with those of \texttt{QArray+}, while \autoref{fig:models} shows representative CSDs computed with the original \texttt{QArray} and with the new \texttt{QArray+} models. In \texttt{QArray+}, all capacitance matrices and tunnel couplings include explicit gate-voltage dependence, allowing the simulator to reproduce device behavior across a wide range of operating conditions. These capabilities are provided through a unified interface, with optional functionality for generating charge-sensor signals and adding realistic noise to synthetic CSDs. The full stack is implemented in JAX, enabling just-in-time compilation and accelerator execution. In particular, the simulator is designed to scale from a single GPU to multi-GPU and multi-node environments. See \autoref{fig: schematic} for a schematic representation of a quantum dot device and \autoref{fig: structure} for an overview of the structure of the software packages. In addition to these three dynamical models, the package provides a fourth simulation head---an integer ground-state (steady-state) solver that deterministically selects the minimum-energy charge configuration at each gate-voltage point, reproducing the steady-state model of the original \texttt{QArray} [used for the steady-state panel of \autoref{fig:models}(a)]---alongside the original \texttt{QArray} model for backward compatibility as seen in Appendix \ref{app:comparison}.

In the following \autoref{sec:models}, we introduce the models used to compute CSDs. Then in \autoref{sec:results} we give examples of the CSDs generated by these models and discuss their features. Finally, in \autoref{sec:benchmarking} we benchmark the models on CPUs, on a single GPU and on multiple GPUs.

\section{Models for simulating charge stability diagrams} \label{sec:models}

In this section, we describe the three dynamical models used to compute the charge stability diagrams in \texttt{QArray+}.

\subsection{Stochastic Capacitance Model}\label{sec:stochastic}
To model the non-equilibrium charge dynamics responsible for latching behavior, we evolve the charge configuration $\vec{n}$ of a quantum-dot array in discrete time steps, each corresponding to the integration time of the sensor measurement (i.e., one pixel of a charge-stability diagram). As a first approximation, we assume that within each integration window, at most one charge transition may occur, consistent with the experimentally relevant regime in which tunnel rates are low compared with the sampling bandwidth. As shown in \autoref{sec:open}, a classical Monte Carlo model of exactly this structure emerges from the open-quantum-system treatment in the limit of vanishing interdot tunnel couplings, grounding the stochastic approach within the hierarchy of timescales that characterizes charge dynamics in quantum dot arrays.

At the start of each time step, the algorithm enumerates all elementary charge-transfer processes:
\begin{enumerate}
    \item \textbf{Loading/Unloading:} an electron tunnels from a reservoir into or out of a dot~$i$, changing $n_i \to n_i \pm 1$;
    \item \textbf{Interdot tunneling:} an electron tunnels between neighboring dots $i$ and $j$, conserving total electron number.
\end{enumerate}
For each possible transition, the change in electrostatic energy $\Delta U$ is computed according to the capacitance model for quantum dots \cite{RevModPhys.75.1, van_Straaten_2024, murphy2025rfsquadradiofrequencysimulatorquantum, Krzywda_2025, Gualtieri_2025}, such that 
\begin{align}
    U(\vec{n}) 
      &= \frac{1}{2}\,\vec{Q}^{\mathsf{T}} \mathbf{C}_{dd}^{-1} \vec{Q}, \label{eq:U}
      &&\text{where } 
         \vec{Q} := |e|\vec{n} - \mathbf{C}_{dg}\vec{V}_g,
\end{align} 
where matrices $\mathbf{C}_{dd}$ and $\mathbf{C}_{dg}$ denote the dot-dot and gate-dot capacitances, respectively, and may be functions of gate voltage. As in the constant-interaction picture, \autoref{eq:U} contains no quantized orbital level spacing. Although occupation-dependent orbital energies are not explicitly modeled, their smooth gate-voltage dependence can be partially emulated through the voltage-dependent capacitance matrices~\cite{Hanson_review}. That is, in general,
\begin{align}
    \mathbf{C}_{dd} := \mathbf{C}_{dd}(\vec{V}_g) \text{ and } \mathbf{C}_{dg} := \mathbf{C}_{dg}(\vec{V}_g).
\end{align}
In the following, we describe how tunneling rates are calculated for loading, unloading, and interdot transitions. 

\paragraph{Lead-mediated loading and unloading -}
Electron exchange between dot~$i$ and its reservoir occurs in the weak-tunneling regime and is therefore governed by Fermi--Dirac statistics. The loading rates are $\Gamma^{\mathrm{loading}}_i = \Gamma_i(\vec{V}_g)/(1 + \exp(\beta \Delta U_i))$, and the corresponding unloading rates are $\Gamma^{\mathrm{unloading}}_i = \Gamma_i(\vec{V}_g)/(1 + \exp(-\beta \Delta U_i))$, where $\Delta U_i$ denotes the change in electrostatic energy associated with the $i$th loading transition, $\Gamma_i$ is a (possibly gate-dependent) tunneling rate specified by the user, and $\beta = (k_B T)^{-1}$ is the inverse temperature. 

\paragraph{Interdot transitions -}

For interdot tunneling, the user specifies a rank-3 tensor $\Gamma^{\mathrm{inter}}_{s,ij}$, which parameterizes the transition rates between quantum dots $i$ and $j$. The index $s \in \{0,1\}$ indicates whether Pauli spin blockade (PSB) is relevant for the transition: $\Gamma^{\mathrm{inter}}_{0,ij}$ corresponds to transitions that are not subject to PSB, whereas $\Gamma^{\mathrm{inter}}_{1,ij}$ corresponds to transitions that are subject to PSB. For a transition between dots $i$ and $j$, the PSB rate is applied when both dots contain an odd number of charges prior to the transition. Similarly to the lead loading, we modulate the rate of interdot tunneling with the Fermi function to account for temperature effects, such that $\Gamma^{\mathrm{inter}}_{ij} \to \Gamma^{\mathrm{inter}}_{ij} / (1 + \exp(\beta \Delta U_{ij}))$, where $\Delta U_{ij}$ denotes the change in electrostatic energy upon an interdot transition between dots $i$ and $j$, computed according to \autoref{eq:U}. This Fermi factor is a bounded interpolation: it preserves the detailed-balance ratio $e^{-\beta \Delta U_{ij}}$ between forward and backward transitions---and hence the correct equilibrium populations---while remaining finite at the interdot resonance, where the bosonic (phonon) factor of the open-system model (\autoref{sec:open}) is instead resonantly enhanced, a divergence that is physically cut off by the phonon spectral density. This treatment is also an approximation in that it does not account for coherent tunneling between the dots; coherent tunneling is captured by the Lindblad-based open-system simulator discussed later.

\paragraph{Stochastic charge state update -}

Having defined the transition rates, we now describe how the charge state is
updated stochastically to reflect charge transfer events. For an integration
window of duration $\tau$, the probability that a transition with rate $\Gamma_k$ occurs within this window is given by
\begin{equation}
	p_k = 1 - \exp(-\Gamma_k \tau ).
\end{equation}

\noindent To update the charge configuration, we perform a stochastic (dynamic Monte Carlo)
update based on the transition probabilities defined above. The update procedure
is as follows:
\begin{enumerate}
    \item all allowed transitions are traversed in a random order;
    \item for each transition, a Bernoulli trial with success probability $p_k$
    is performed;
    \item the first transition for which the trial succeeds is executed
    immediately; if no trial succeeds, the charge configuration remains
    unchanged.
\end{enumerate}
The procedure is then repeated for the next pixel of the charge stability
diagram, starting from the updated charge configuration. 

This single-transition-per-time-step scheme is accurate in the slow-tunneling
regime, where $\Gamma_k \tau \ll 1$ for all relevant transitions. The scheme
reproduces the total jump probability exactly, since
$\prod_k (1 - p_k) = e^{-\tau \sum_k \Gamma_k}$, but the random traversal order
biases the selection among competing channels relative to the exact branching
ratios $\Gamma_k / \sum_j \Gamma_j$ at second order in $\Gamma_k \tau$
[cf.\ \autoref{eq:classical_branching}]. When
multiple transitions within a single integration time must be resolved, or when
$\Gamma_k \tau \ll 1$ is not satisfied, the
method can be generalized by subdividing the integration window into $N_r$
sub-intervals of duration $\tau/N_r$ and applying the update rule
recursively within each sub-interval, which suppresses this bias accordingly.

\paragraph{Computational complexity -}
For a system consisting of $n_{\mathrm{dot}}$ dots there are only $N_r n_{\mathrm{dot}}(n_{\mathrm{dot}} + 1)$ charge transitions to evaluate, so the computational complexity of this approach scales polynomially with the number of dots. This polynomial scaling represents a dramatic improvement over steady-state solvers, which typically must evaluate the electrostatic energy of an exponentially growing number of charge states.  See Appendix~\ref{app:complexity} for a complete discussion of the complexity.

\subsection{Tunnel coupling}

The capacitance model used in the original \texttt{QArray} and in the previous model treats each charge
configuration $\vec n \in \mathbb{Z}_{\ge 0}^{n_{\mathrm{dot}}}$ as a classical
state with a well-defined electrostatic energy. However, in many experimentally
relevant regimes, finite interdot tunnel coupling hybridizes charge configurations close to degeneracy, producing avoided crossings and fractional
charge expectation values~\cite{van_Riggelen_Doelman_2024}.

To address this limitation, we extend the capacitance model into a spinless Hubbard model, with a coherent tunneling Hamiltonian defined directly in the charge (occupation-number) basis, similar to that introduced in \cite{murphy2025rfsquadradiofrequencysimulatorquantum}. Then we compute CSDs by assuming that, for each gate voltage, the system adopts the lowest-energy eigenstate of this new charge-basis Hamiltonian.

\paragraph{Charge-basis Hamiltonian -}
We work in the occupation-number basis
$\{ \ket{\vec n} \} = \{ \ket{n_1,\dots,n_{n_{\mathrm{dot}}}} \}$ with
$n_i \in \{0,1,\dots\}$.
The diagonal (electrostatic) contribution is taken to be
\begin{equation}
\hat H_F(\vec V_g) \;=\; \sum_{\vec n} F(\vec n;\vec V_g)\,\ket{\vec n}\!\bra{\vec n},
\end{equation}
where $F(\vec n;\vec V_g)$ is the electrostatic energy obtained from the capacitance model, \autoref{eq:U}. When the capacitances are themselves gate-voltage dependent, $F$ should be read as a per-pixel electrostatic energy landscape rather than a thermodynamic free energy.
In this formalism, overall constant prefactors (e.g.\ $|e|^2/2$) are absorbed
into the choice of energy units; only relative energies enter the ground-state
problem.

Coherent interdot tunneling is modeled by single-mode hopping terms written in
terms of charge ladder operators $\hat a_j,\hat a_i^\dagger$ at each site acting as
$\hat a_i\ket{\dots,n_i,\dots}=\sqrt{n_i}\ket{\dots,n_i-1,\dots}$ and
$\hat a_i^\dagger\ket{\dots,n_i,\dots}=\sqrt{n_i+1}\ket{\dots,n_i+1,\dots}$.
With an optional, user-specified gate-dependent tunnel matrix $\mathbf{t}(\vec V_g)$,
the tunneling Hamiltonian is
\begin{equation}
\hat H_t(\vec V_g) \;=\; -\sum_{i\neq j} t_{ij}(\vec V_g)\,\hat a_j^\dagger \hat a_i .
\label{eq:Ht}
\end{equation}
Hermiticity of $\hat H_t$ requires $t_{ij} = t_{ji}^{*}$; in practice, the tunnel couplings are
taken to be real and symmetric.
In the charge basis, the only nonzero off-diagonal matrix elements connect states
that differ by moving one electron from dot $i$ to dot $j$:
\begin{equation}
\bra{\vec n'} \hat H_t \ket{\vec n}
=
-t_{ij}(\vec V_g)\,\sqrt{n_i (n_j+1)}
\quad\text{if}\quad
\vec n' = \vec n - \hat e_i + \hat e_j,
\end{equation}
and vanish otherwise, where $\hat e_i$ denotes the $i$th standard basis vector. Note that the
charge ladder operators endow the hopping element with a soft-core bosonic enhancement
$\sqrt{n_i (n_j+1)}$ relative to the bare element $-t_{ij}$ of a spinless fermionic Hubbard
model with $n_i \in \{0,1\}$~\cite{Krzywda_2025, murphy2025rfsquadradiofrequencysimulatorquantum}; the two conventions
coincide for transitions between empty and singly occupied dots and differ only when multiply
occupied dots participate. The full
Hamiltonian is then
\begin{equation}
\hat H(\vec V_g) \;=\; \hat H_F(\vec V_g) + \hat H_t(\vec V_g).
\end{equation}

\paragraph{Tunnel-coupled ground state and charge expectation values -}
For each gate-voltage point $\vec V_g$, the tunnel-coupled ground state
$\ket{\psi_0(\vec V_g)}$ is obtained by solving the eigenproblem
$\hat H(\vec V_g)\ket{\psi_0}=E_0\ket{\psi_0}$ and selecting the lowest eigenvalue.
The simulator reports the \emph{expected} dot occupations
\begin{equation}
\langle n_k \rangle(\vec V_g)
=
\bra{\psi_0(\vec V_g)} \hat n_k \ket{\psi_0(\vec V_g)}
=
\sum_{\alpha} |\psi_{0,\alpha}(\vec V_g)|^2\, n_{\alpha,k},
\end{equation}
where $\alpha$ indexes the charge-basis states included in the calculation and
$n_{\alpha,k}$ denotes the occupation of dot $k$ in basis state $\alpha$.

\paragraph{Efficient truncation of the charge basis -}
A naive implementation would require diagonalizing $\hat H$ in the full charge basis,
whose dimension grows exponentially with $n_{\mathrm{dot}}$ and the maximum charge
per dot. Instead, \texttt{QArray+} constructs a \emph{truncated} charge basis of
fixed size $n_{\mathrm{truncate}}$ that captures the low-energy manifold relevant
near the ground state.

Concretely, for each $\vec V_g$ we first compute a continuous ``open-charge''
solution $\vec n_{\mathrm{cont}}(\vec V_g)$---the minimizer of \autoref{eq:U} with the
integrality constraint on $\vec n$ relaxed---by solving the corresponding convex
quadratic program.
We then define an integer baseline $\vec n_{\mathrm{base}}$ by taking the element-wise
floor of $\vec n_{\mathrm{cont}}$ and generate a mixed-radix set of
candidate integer configurations of the form
$\vec n = \vec n_{\mathrm{base}} + \vec\delta$,
where $\delta_i$ is drawn from a small dot-dependent set. In the default heuristic
used in the code, dots that couple most strongly to the gates are assigned a
larger neighborhood $\delta_i \in \{-1,0,1,2\}$, more weakly coupled dots are
assigned $\delta_i \in \{0,1\}$, and all remaining dots are fixed by rounding;
these asymmetric default neighborhoods are a heuristic choice.
The energy $F(\vec n;\vec V_g)$ of each candidate is evaluated efficiently
using a Cholesky factorization of $\mathbf{C}_{dd}$, and the
$n_{\mathrm{truncate}}$ lowest-energy candidates define the truncated basis.

Finally, $\hat H(\vec V_g)$ is assembled in this truncated basis and the ground
state is computed either by a dense eigensolve (for small $n_{\mathrm{truncate}}$)
or by a sparse Lanczos iteration (for larger truncated dimensions). This yields an
accurate approximation to the tunnel-coupled ground state while avoiding the
exponential scaling of the full basis.

\paragraph{Computational complexity -}

Consider a system of $n_{\mathrm{dot}}$ dots, and let
$n_{\mathrm{truncate}}$ denote the truncated-basis dimension used for the eigensolve. For
each gate-voltage point $\vec V_g$, the solver consists of: (i) generating a candidate set
$\mathcal{C}(\vec V_g)$ of integer charge configurations near the relaxed solution
$\vec n_{\mathrm{cont}}(\vec V_g)$, (ii) evaluating $F(\vec n;\vec V_g)$ for all
$\vec n\in\mathcal{C}$ and selecting the $n_{\mathrm{truncate}}$ lowest-energy candidates,
(iii) assembling the truncated Hamiltonian, and (iv) computing its ground state.

Let $N_{\mathrm{cand}}:=|\mathcal{C}(\vec V_g)|$ denote the number of candidate configurations
and $E_t$ the number of nonzero interdot
tunnel couplings $t_{ij}$ (i.e.\ edges in the tunnel graph). Given a Cholesky factorization of
$\mathbf{C}_{dd}$, each energy evaluation $F(\vec n;\vec V_g)$ can be performed using two
triangular solves and a dot product, scaling as $\mathcal{O}(n_{\mathrm{dot}}^{2})$.
Thus candidate scoring scales as $\mathcal{O}(N_{\mathrm{cand}}\,n_{\mathrm{dot}}^{2})$ per voltage point
(up to an additional $\mathcal{O}(n_{\mathrm{dot}}^{3})$ per point if $\mathbf{C}_{dd}$ must be refactorized
due to gate-voltage dependence).

The truncated Hamiltonian is sparse because $\hat H_t$ connects only charge states that differ by a
single hop; the number of nonzero off-diagonal elements scales as
$\mathrm{nnz}(\hat H_t)=\mathcal{O}(n_{\mathrm{truncate}}\,E_t)$.
Computing the ground state by dense diagonalization costs $\mathcal{O}(n_{\mathrm{truncate}}^{3})$,
whereas a Krylov/Lanczos method costs $\mathcal{O}(n_{\mathrm{iter}}\;\mathrm{nnz}(\hat H))$ with
$n_{\mathrm{iter}}$ iterations. Overall, the full scan therefore scales as
\[
\mathcal{O}\!\Bigl(
N^{2}\bigl[
N_{\mathrm{cand}}\,n_{\mathrm{dot}}^{2}
+ n_{\mathrm{truncate}}\,E_t
+ \mathrm{eig}(n_{\mathrm{truncate}})
\bigr]\Bigr),
\]
where $N$ is the linear dimension of the $N \times N$ gate-voltage grid and
$\mathrm{eig}(n_{\mathrm{truncate}})\in\{\mathcal{O}(n_{\mathrm{truncate}}^{3}),
\mathcal{O}(n_{\mathrm{iter}}\,n_{\mathrm{truncate}}\,E_t)\}$ depending on the eigensolver.
Crucially, by keeping $n_{\mathrm{truncate}}$ (and in practice $N_{\mathrm{cand}}$) fixed and small,
the computation is controlled by the truncated manifold rather than by the full charge basis,
whose dimension grows exponentially with $n_{\mathrm{dot}}$.

\subsection{Open quantum system}\label{sec:open}

The stochastic charge-transition model described above treats charge dynamics as a classical Markov jump process between integer occupation states, capturing latching but not hybridization, while the spinless Hubbard model captures hybridization but not latching~\cite{daley2014quantum,plenio1998quantum}. To simulate both effects simultaneously, we treat the quantum dot array as an open quantum system. To model the charge stability diagram while capturing both coherent hybridization and stochastic charging and relaxation dynamics, we employ an instantaneous-eigenbasis stochastic Schr\"odinger equation (SSE) approach. Unlike standard Fock-basis quantum-jump models, which project the system onto unhybridized charge configurations at each jump, this formulation allows coherent superpositions to survive at avoided crossings between dissipative events~\cite{breuer2002theory}.

\paragraph{Charge basis, Hamiltonian, and master equation -}
We again work in the charge basis $\{\ket{\vec n}\}$, but now explicitly truncate the
local occupations to $n_i \in \{0,1,\dots,n_{\max}\}$, yielding a Hilbert-space
dimension $S=(n_{\max}+1)^{n_{\mathrm{dot}}}$.
For each gate-voltage point $\vec V_g$, we construct the same tunnel-coupled Hamiltonian
\begin{equation}
\hat H(\vec V_g)=\hat H_F(\vec V_g)+\hat H_t(\vec V_g),
\end{equation}
with $\hat H_F$ diagonal in the charge basis and $\hat H_t$ given by the hopping term in
\autoref{eq:Ht}. In this open-system setting, $\hat H$ generates coherent evolution between
charge configurations that are connected by interdot tunneling. The system evolves under the
Lindblad master equation (setting $\hbar = 1$)
\begin{equation}
\dot\rho = -i\,[\hat H(\vec V_g),\rho] + \mathcal{D}(\rho),
\label{eq:master_equation}
\end{equation}
where the dissipator $\mathcal{D}$ comprises contributions from a bosonic phonon bath and from
the fermionic leads:
\begin{equation}
\mathcal{D}(\rho) = \mathcal{D}_{\mathrm{phon}}(\rho) + \mathcal{D}_{\mathrm{lead}}(\rho).
\label{eq:dissipator_split}
\end{equation}

\paragraph{Bosonic phonon bath -}
Incoherent interactions with phonons induce transitions between the instantaneous eigenstates
$\ket{\psi_\alpha}$ of $\hat H(\vec V_g)$, with energies $E_\alpha$, while preserving the total
electron number $\hat N = \sum_{i} \hat n_i$. Because $\hat H$ conserves $\hat N$, every
eigenstate carries a definite total charge $N_\alpha$, and these processes drive relaxation and
dephasing within each fixed-charge sector:
\begin{equation}
\mathcal{D}_{\mathrm{phon}}(\rho)
= \sum_{\alpha \neq \beta} D\!\left[\sqrt{\kappa_{\alpha\beta}}\,\ket{\psi_\alpha}\!\bra{\psi_\beta}\right]\!(\rho),
\label{eq:phonon_dissipator}
\end{equation}
where $D[\hat L](\rho) = \hat L\rho \hat L^\dagger - \tfrac{1}{2}\{\hat L^\dagger \hat L,\rho\}$
is the Lindblad dissipator and the sum runs over pairs of eigenstates within the same
total-charge sector ($N_\alpha = N_\beta$). The transition rates are those of a thermal bosonic
bath,
\begin{equation}
\kappa_{\alpha\beta}
= \begin{cases}
\Gamma_{\mathrm{phon}}\bigl[1 + n_B(|\Delta E_{\alpha\beta}|)\bigr], & \Delta E_{\alpha\beta} < 0,\\[3pt]
\Gamma_{\mathrm{phon}}\, n_B(\Delta E_{\alpha\beta}), & \Delta E_{\alpha\beta} > 0,
\end{cases}
\label{eq:phonon_rates}
\end{equation}
where $\Delta E_{\alpha\beta} = E_\alpha - E_\beta$, $\Gamma_{\mathrm{phon}}$ is the phonon
coupling strength, and $n_B(E) = (e^{E/(k_B T)} - 1)^{-1}$ is the Bose--Einstein occupation:
downward transitions proceed at the stimulated-plus-spontaneous emission rate and upward
transitions at the absorption rate. The rates obey the quantum detailed-balance relation for a
thermal bath,
$\kappa_{\alpha\beta}/\kappa_{\beta\alpha} = e^{-\Delta E_{\alpha\beta}/(k_B T)}$, which follows
from $[1 + n_B(E)]/n_B(E) = e^{E/(k_B T)}$.
This bath acts as a dephasing mechanism, decohering superpositions on a timescale
$\sim 1/\Gamma_{\mathrm{phon}}$ while allowing relaxation within the fixed-charge manifold.

\paragraph{Fermionic lead coupling -}
Charge carriers tunnel between the dots and external fermionic reservoirs, mediating transitions
between charge states $\vec n$ and $\vec n \pm \hat e_i$, where $\hat e_i$ is the $i$th standard
basis vector. The corresponding dissipator is
\begin{equation}
\mathcal{D}_{\mathrm{lead}}(\rho)
= \sum_{i,\pm,\vec n} D\!\left[\sqrt{\kappa_{i,\pm}(\vec n)}\,\ket{\vec n \pm \hat e_i}\!\bra{\vec n}\right]\!(\rho),
\label{eq:lead_dissipator}
\end{equation}
where the sum retains only moves that respect the truncation $0 \le n_i \le n_{\max}$. The
transition rate is thermally regulated by the Fermi factor
\begin{equation}
\kappa_{i,\pm}(\vec n)
= \frac{\Gamma_i}{1 + \exp\!\bigl[\Delta E_{i,\pm}(\vec n)/(k_B T)\bigr]},
\label{eq:lead_rates}
\end{equation}
where $\Gamma_i$ is the dot--lead coupling strength and
$\Delta E_{i,\pm}(\vec n) = U(\vec n \pm \hat e_i;\vec V_g) - U(\vec n;\vec V_g)$ is the
electrostatic energy cost of adding ($+$) or removing ($-$) an electron on dot $i$, computed
according to \autoref{eq:U}. These rates coincide with the lead-mediated loading and unloading
rates used in the stochastic capacitance model.

\paragraph{Hierarchy of timescales -}
Physical devices exhibit a clear separation of timescales,
\begin{equation}
\Delta_0 \gg \Gamma_{\mathrm{phon}} \gg \Gamma_{\mathrm{lead}},
\label{eq:timescale_hierarchy}
\end{equation}
where $\Delta_0 \sim \max|\mathrm{spec}(\hat H)|$ sets the overall spectral scale of
$\hat H$ and $\Gamma_{\mathrm{lead}}$ denotes the typical magnitude of the dot--lead
couplings $\Gamma_i$. This ordering partitions the dynamics into three distinct regimes. On the
\emph{coherent timescale} $\tau_{\mathrm{coh}} \sim 1/\Delta_0$, the system exhibits nearly
coherent evolution between electronic eigenstates, with dissipation negligible. On the
\emph{phonon timescale} $\tau_{\mathrm{phon}} \sim 1/\Gamma_{\mathrm{phon}}$, the phonon bath
gradually decoheres oscillations through incoherent transitions that preserve charge. Finally,
on the \emph{lead timescale} $\tau_{\mathrm{lead}} \sim 1/\Gamma_{\mathrm{lead}}$, charge
transfer between dots and reservoirs modulates the total charge state.

When measuring a charge stability diagram, we assume that the gate voltages change
instantaneously between adjacent pixels and that the charge-sensor signal is integrated over a
time interval $\tau_{\mathrm{int}}$, corresponding to one pixel. We simulate the experimentally
realistic regime
\begin{equation}
\tau_{\mathrm{phon}} \ll \tau_{\mathrm{int}} \lesssim \tau_{\mathrm{lead}},
\label{eq:measurement_regime}
\end{equation}
ensuring that the integration time is much longer than phonon relaxation timescales but
comparable to the lead-induced decoherence timescale; for reference, typical integration times
are on the microsecond timescale. Under these conditions, coherences between eigenstates are
damped out within each pixel, and we treat the system as evolving stochastically among the
Hamiltonian's instantaneous eigenstates. Note that because these eigenstates are themselves
hybridized superpositions of charge configurations, interdot charge coherence is retained within
each eigenstate. For our charge-stability-diagram simulations, we initialize the first pixel in
the instantaneous Hamiltonian's ground state and then propagate this state using the stochastic
wavefunction propagation described next.

\paragraph{Stochastic wavefunction propagation -}
The dynamics of an open quantum system subject to inelastic transitions and decoherence are
generally described by a master equation governing the density matrix $\rho(t)$, but solving
\autoref{eq:master_equation} directly becomes computationally expensive for systems with many
eigenstates. An alternative approach is to represent the evolution stochastically using the
quantum-jump method (also known as Monte Carlo wavefunction simulation): pure-state trajectories
evolve under a non-Hermitian effective Hamiltonian, interspersed with stochastic collapse events
representing inelastic transitions. When averaged over many trajectories, this approach
reproduces the master equation's predictions while remaining numerically efficient. In our case,
however, we are not interested in the ensemble average, as charge stability diagrams are
typically measured in a small number of shots; a single trajectory directly mimics an individual
experimental scan.

Under the quantum-jump protocol, the state $\ket{\Psi(t)}$ evolves over a timestep
$\Delta t = \tau_{\mathrm{int}}$ according to the following procedure. We first expand the state
in the instantaneous eigenbasis of $\hat H(\vec V_g)$:
\begin{equation}
\ket{\Psi(t)} = \sum_\beta c_\beta(t)\,\ket{\psi_\beta(t)}.
\label{eq:eigenbasis_expansion}
\end{equation}
The non-Hermitian effective Hamiltonian
\begin{equation}
\hat H_{\mathrm{eff}} = \hat H - \frac{i}{2}\sum_\alpha \Lambda_\alpha \ket{\psi_\alpha}\!\bra{\psi_\alpha}
\label{eq:Heff}
\end{equation}
generates the unnormalized coefficients over the timestep $\Delta t$:
\begin{equation}
c_\beta(t + \Delta t) = c_\beta(t)\,
\exp\!\left[\left(-i E_\beta - \frac{1}{2}\Lambda_\beta\right)\Delta t\right],
\label{eq:coefficient_evolution}
\end{equation}
where $\Lambda_\beta = \sum_{\alpha \neq \beta} R_{\alpha\beta}$ is the total escape rate from
eigenstate $\ket{\psi_\beta}$, with $R_{\alpha\beta}$ a matrix encoding the transition rate from
eigenstate $\ket{\psi_\beta}$ to eigenstate $\ket{\psi_\alpha}$; we outline how the matrix
$R_{\alpha\beta}$ is populated below.
The imaginary part of the exponent encodes the coherent energy evolution, while the real part
accounts for exponential decay due to dissipation---this is the non-Hermitian evolution that
underlies the master-equation approach. Because the gate voltages are constant within a pixel,
the eigenbasis is fixed during each timestep, and \autoref{eq:coefficient_evolution} follows
exactly from \autoref{eq:Heff}.

The probability that at least one quantum jump occurs during the timestep is
\begin{equation}
P_{\mathrm{jump}} = 1 - \sum_\beta |c_\beta(t+\Delta t)|^2.
\label{eq:pjump}
\end{equation}
At each step, a uniform random variable $\xi \in [0,1)$ determines which of two scenarios
occurs.

\textbf{Case 1: No jump} ($\xi \ge P_{\mathrm{jump}}$). The state undergoes continuous evolution
under $\hat H_{\mathrm{eff}}$, and the coefficients are renormalized to restore unit norm:
\begin{equation}
\ket{\Psi(t+\Delta t)}
= \frac{\sum_\beta c_\beta(t+\Delta t)\,\ket{\psi_\beta(t+\Delta t)}}{\sqrt{1 - P_{\mathrm{jump}}}}.
\label{eq:no_jump}
\end{equation}

\textbf{Case 2: Quantum jump} ($\xi < P_{\mathrm{jump}}$). A stochastic collapse occurs,
representing an inelastic transition. The source eigenstate $\ket{\psi_j}$ is sampled according
to the probability that the norm loss originated from that state,
\begin{equation}
P(\mathrm{source} = j)
= \frac{|c_j(t)|^2\left(1 - e^{-\Lambda_j \Delta t}\right)}{P_{\mathrm{jump}}},
\label{eq:jump_source}
\end{equation}
which reduces to $P(\mathrm{source} = j) \propto |c_j(t)|^2\,\Lambda_j$ to first order in
$\Delta t$. Given the source, the target eigenstate $\ket{\psi_i}$ is randomly selected via the
branching ratio
\begin{equation}
P(\mathrm{target} = i \mid \mathrm{source} = j) = \frac{R_{ij}}{\Lambda_j}.
\label{eq:branching_ratio}
\end{equation}
The state then collapses to the target eigenstate:
\begin{equation}
\ket{\Psi(t+\Delta t)} = \ket{\psi_i(t+\Delta t)}.
\label{eq:collapse}
\end{equation}
Note that, mirroring the single-transition-per-time-step scheme of the stochastic capacitance
model, at most one quantum jump is applied per evolution substep (by default, a single substep
spans the whole integration window). In parallel mode (introduced
below), each row is re-initialized in the instantaneous ground state, so this one-jump update
acts on a state that has already relaxed. In continuous mode, where the trajectory is carried
between pixels, the regime of \autoref{eq:measurement_regime} implies
$\Gamma_{\mathrm{phon}} \tau_{\mathrm{int}} \gg 1$, and the integration window should therefore
be subdivided into $N_r \gtrsim \Gamma_{\mathrm{phon}} \tau_{\mathrm{int}}$ substeps---in analogy
with the recursion depth of the stochastic capacitance model---so that the jump probability per
substep remains small.

The simulated charge signal for a given pixel is then obtained by computing the expectation
value of the charge occupation vector,
$\bra{\Psi(t+\Delta t)}\hat{\vec n}\ket{\Psi(t+\Delta t)}$, where the components of
$\hat{\vec n} = (\hat n_1,\dots,\hat n_{n_{\mathrm{dot}}})$ correspond to the number operators of
the individual quantum dots. Because the eigenstates are hybridized superpositions of charge
configurations, this signal captures the fractional charge expectation values responsible for
avoided crossings, while the stochastic jumps reproduce latching dynamics; in the limit of
vanishing $t_{ij}$, the diagrams reduce to the classical latching picture.

\paragraph{Scan modes -}
When computing a two-dimensional charge stability diagram, the gate voltages are typically swept
in a raster pattern. To simulate this, we provide two distinct computational modes. In
\emph{continuous} mode, the entire diagram evolves sequentially: the final charge configuration
of one pixel serves as the initial state for the subsequent pixel, preserving the stochastic
history of the system across the entire scan, including the flyback between rows. Alternatively,
in \emph{parallel} mode, we assume the system fully relaxes during the raster flyback or the
subsequent wait time before starting the row, such that each new row is initialized in the
instantaneous ground state. This assumption removes the physical dependence between rows,
rendering the calculation embarrassingly parallel and significantly reducing computation time.

\paragraph{Dissipative jump rates in the eigenbasis -}
To simulate the stochastic trajectories described above, we must compute the transition-rate
matrix $R_{\alpha\beta}$, which defines the probability per unit time of a quantum jump from the
instantaneous source eigenstate $\ket{\psi_\beta}$ to the target eigenstate $\ket{\psi_\alpha}$.
The total transition rate is the sum of the independent contributions from the phonon bath and
the fermionic leads:
\begin{equation}
R_{\alpha\beta} = R^{\mathrm{phon}}_{\alpha\beta} + R^{\mathrm{lead}}_{\alpha\beta}.
\label{eq:total_rates}
\end{equation}
Crucially, by constructing the jump rates directly between the instantaneous energy eigenstates,
we are applying the secular approximation, thereby explicitly dropping the jump-induced
coherences (off-diagonal terms in the density matrix). This approximation is physically
justified by the system's separation of timescales
($\Delta_0 \gg \Gamma_{\mathrm{phon}}, \Gamma_{\mathrm{lead}}$), ensuring that the coherent
energy scale strictly dominates and any rapidly oscillating off-diagonal coherences average to
zero. As a result, inelastic collapse events map populations directly to populations in the
instantaneous eigenbasis. We note that $\Delta_0$ characterizes the overall spectral scale; the
locally relevant frequency for a given pair of eigenstates is their splitting
$|E_\alpha - E_\beta|$, and the secular condition therefore fails locally at avoided crossings,
where the gap collapses to $\sim 2 t_{ij}$ and can become comparable to the dissipative rates.
In these narrow regions the secular construction should be regarded as an approximation.

The phonon dissipator $\mathcal{D}_{\mathrm{phon}}$ is already defined via transitions between
system eigenstates. Thus, the phonon transition rate follows directly from the detailed-balance
coefficients of \autoref{eq:phonon_rates},
\begin{equation}
R^{\mathrm{phon}}_{\alpha\beta} = \kappa_{\alpha\beta}
\quad \text{for } N_\alpha = N_\beta \text{ and } \alpha \neq \beta,
\label{eq:phonon_rate_matrix}
\end{equation}
where $N_\alpha$ and $N_\beta$ are the total charges of the target and source eigenstates,
respectively; all other elements vanish, and in particular
$R^{\mathrm{phon}}_{\alpha\alpha} = 0$.

For the lead dissipator $\mathcal{D}_{\mathrm{lead}}$, the jump operators are defined in the
local charge basis $\ket{\vec n}$. To find the eigenstate-to-eigenstate transition rate, we must
project these jump operators into the eigenbasis. Let an instantaneous eigenstate be expanded in
the charge basis as $\ket{\psi_\gamma} = \sum_{\vec n} c^{(\gamma)}_{\vec n} \ket{\vec n}$, where
$c^{(\gamma)}_{\vec n} = \braket{\vec n | \psi_\gamma}$. Each Lindblad jump operator $\hat L$
contributes a transition rate $|\bra{\psi_\alpha} \hat L \ket{\psi_\beta}|^2$ from
$\ket{\psi_\beta}$ to $\ket{\psi_\alpha}$. Summing the contributions of all single-electron lead
jump operators yields
\begin{align}
R^{\mathrm{lead}}_{\alpha\beta}
&= \sum_{i,\pm,\vec n}
\Bigl|\bra{\psi_\alpha}\Bigl(\sqrt{\kappa_{i,\pm}(\vec n)}\,\ket{\vec n \pm \hat e_i}\!\bra{\vec n}\Bigr)\ket{\psi_\beta}\Bigr|^2
\nonumber\\
&= \sum_{i,\pm,\vec n} \kappa_{i,\pm}(\vec n)\,
\bigl|c^{(\alpha)}_{\vec n \pm \hat e_i}\bigr|^2\,\bigl|c^{(\beta)}_{\vec n}\bigr|^2.
\label{eq:lead_rate_matrix}
\end{align}
This result carries a clear physical interpretation: the rate of an electron tunneling into or
out of the dot array is governed by the underlying tunneling rate $\kappa_{i,\pm}(\vec n)$,
weighted by the probability $|c^{(\beta)}_{\vec n}|^2$ that the initial eigenstate is in the
necessary pre-tunneling charge configuration $\ket{\vec n}$, and multiplied by the probability
$|c^{(\alpha)}_{\vec n \pm \hat e_i}|^2$ that the target eigenstate supports the resulting
post-tunneling charge configuration $\ket{\vec n \pm \hat e_i}$. Summing the $(i,\pm,\vec n)$
channels incoherently drops interference between different charge-basis pathways relative to a
single coherent lead operator; this is the natural choice here because the rate
$\kappa_{i,\pm}(\vec n)$ depends explicitly on the pre-jump configuration $\vec n$. Finally, the total escape rate
from a given eigenstate $\ket{\psi_\beta}$ is obtained by summing over all possible target
states $\alpha$:
\begin{equation}
\Lambda_\beta = \sum_{\alpha \neq \beta} R_{\alpha\beta}.
\label{eq:escape_rate}
\end{equation}
This rate directly dictates the non-Hermitian decay and the total jump probability for the
stochastic wavefunction propagation.

\paragraph{Classical limit: stochastic capacitance model -}
In the limit where interdot tunneling vanishes ($t_{ij} = 0$ for all $i,j$), the eigenstates
correspond exactly to the charge Fock states $\ket{\vec n}$. This allows for considerable
computational simplification: the state of the system is fully captured by the discrete charge
vector $\vec n$, and the stochastic wavefunction propagation reduces to a classical Monte Carlo
simulation over the charge space.

In this limit, for a given charge state $\vec n$, the fermionic jump operators defined in
\autoref{eq:lead_dissipator} permit transitions to the $2 n_{\mathrm{dot}}$ charge states
differing by one charge in a single dot, with rates computed according to
\autoref{eq:lead_rates}. Similarly, the bosonic jump operators permit internal charge
rearrangements that conserve the total electron number. Because the number of such
charge-preserving configurations scales exponentially with the number of dots, we restrict our
attention to the $n_{\mathrm{dot}}(n_{\mathrm{dot}} - 1)$ pairwise interdot transitions, in
which a single electron is swapped between two dots while the charge state of all other dots
remains unchanged; the corresponding phonon-induced transition rates are computed according to
\autoref{eq:phonon_rates}.

To evolve the system over an integration window of duration $\tau_{\mathrm{int}}$, we apply a
stochastic jump protocol that directly mirrors the wavefunction jump procedure. First, we
compute the total escape rate from the current charge state $\vec n$ by summing over all allowed
fermionic and bosonic transition rates $R_k$:
\begin{equation}
\Lambda_{\vec n} = \sum_k R_k.
\label{eq:classical_escape}
\end{equation}
In the absence of coherent evolution, the probability that at least one charge-transfer event
occurs during the integration window is simply
\begin{equation}
P_{\mathrm{jump}} = 1 - \exp(-\Lambda_{\vec n}\,\tau_{\mathrm{int}}).
\label{eq:classical_pjump}
\end{equation}
A uniform random variable $\xi \in [0,1)$ determines the outcome for the given pixel. If
$\xi \ge P_{\mathrm{jump}}$, no jump occurs: the system remains in the current state and the
charge configuration $\vec n$ is unchanged. If $\xi < P_{\mathrm{jump}}$, a charge-transfer
event occurs, and the specific transition $k$ is randomly selected from the set of allowed
transitions according to its branching ratio:
\begin{equation}
P(\mathrm{target} = k \mid \mathrm{jump}) = \frac{R_k}{\Lambda_{\vec n}}.
\label{eq:classical_branching}
\end{equation}
The charge configuration is immediately updated to the charge state associated with transition
$k$. This procedure is then repeated for the next timestep of the charge stability diagram,
starting from the resulting charge configuration. As with the stochastic wavefunction
propagation, we implement continuous and parallel modes to compute the scan. While this method
sacrifices the ability to capture the coherent hybridization of charge states, it is
considerably more computationally efficient, scaling polynomially rather than exponentially with
the number of dots. This limit recovers the structure of the stochastic capacitance model
introduced in \autoref{sec:stochastic}, with the interdot rates now inherited from the
phonon bath rather than specified directly by the user. The two interdot-rate conventions share
the same detailed-balance ratio, and hence the same equilibrium populations; they differ only in
the lineshape of thermally broadened interdot transitions, where the Fermi factor of
\autoref{sec:stochastic} saturates while the bosonic rate of \autoref{eq:phonon_rates} is
resonantly enhanced.

\paragraph{Steady-state limit: spinless Hubbard model -}
We also consider the complementary regime of long integration times, in which the integration
window exceeds even the lead timescale, such that
\begin{equation}
\tau_{\mathrm{phon}} \ll \tau_{\mathrm{lead}} \ll \tau_{\mathrm{int}}.
\label{eq:steady_state_regime}
\end{equation}
Under these conditions, the quantum dot array fully relaxes at each gate-voltage configuration
and, for temperatures small compared to the relevant energy gaps, settles into its instantaneous
ground state. Consequently, every pixel of the charge stability diagram is
physically independent and can be computed in parallel. We emphasize that
\autoref{eq:measurement_regime} and \autoref{eq:steady_state_regime} describe two distinct
operating regimes of the same model---fast scans that exhibit latching versus slow scans that
equilibrate---selected by the experimental integration time, rather than competing assumptions. In this limit, the open-system model
reduces to the tunnel-coupled ground-state (spinless Hubbard) treatment described above,
including the truncated-basis construction that avoids diagonalizing the Hamiltonian in the
exponentially large full charge basis.

\paragraph{Computational complexity -}
For the general open-system evolution, the dominant cost is governed by the Hilbert-space dimension
$S=(n_{\max}+1)^{n_{\mathrm{dot}}}$: each gate-voltage point requires the eigendecomposition of
$\hat H(\vec V_g)$ and the assembly of the rate matrix $R_{\alpha\beta}$, both of which scale
polynomially in $S$. The classical limit above avoids this cost entirely, which is why the
stochastic capacitance model remains tractable for large arrays.

\section{Results} \label{sec:results}

\subsection{Stochastic Capacitance Model}

\begin{figure*}
    \centering
    \includegraphics{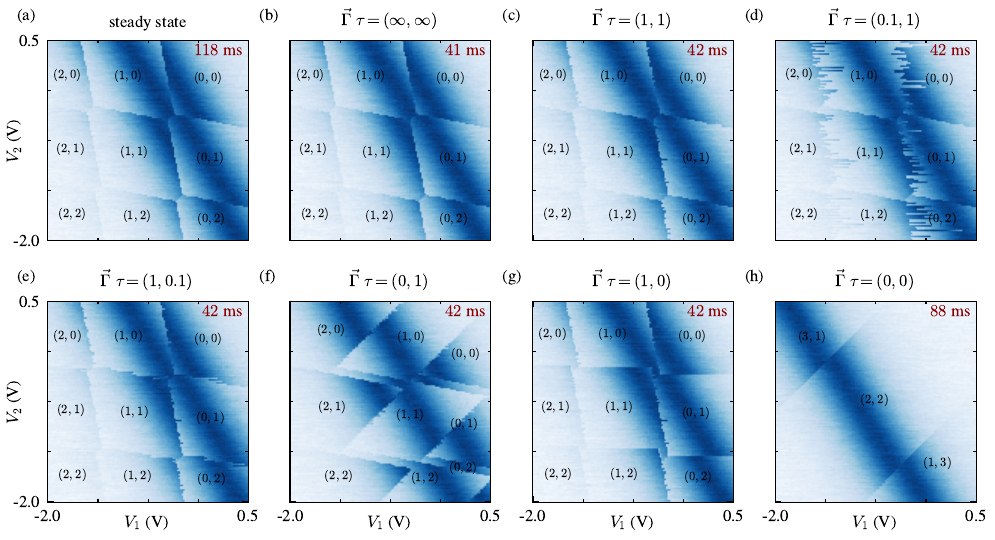}
    \caption{\textbf{Hole-based double-dot charge stability diagrams, each computed on a 
    $200 \times 200$ gate-voltage grid, with the compute time indicated in the 
    top-right corner and the charge state overlaid.} 
    (a) Steady-state solution obtained using the original \texttt{QArray} algorithm. 
    (b) Charge stability diagram computed with the \texttt{QArray+} stochastic solver in the limit of infinite tunnel rates and zero lead temperature, such that the system deterministically adopts the lowest-energy charge configuration.
    (c--g) Latched charge stability diagrams for finite tunnel rates. It is assumed 
    that sufficient time elapses between the end of one row and the start of the next for the system to relax to the ground charge state at the beginning of each row, allowing each row to be computed in parallel. 
    (h) Charge stability diagram for the fully isolated case, where the dots are 
    decoupled from the charge reservoirs, so that only interdot transitions are present. The user specifies the initial charge state in the lower-left corner, and this state is propagated stochastically across the diagram; in this regime, parallelization is not possible. All simulations use $N_r = 1$ and include a charge-sensor readout with $1/f$ noise.}
    \label{fig:double_dot}
\end{figure*}

Figure~\ref{fig:double_dot} illustrates a series of hole-based double-dot charge stability diagrams computed using both the original \texttt{QArray} algorithm and the newly developed \texttt{QArray+} approach. Each diagram is evaluated on a $200 \times 200$ gate-voltage grid, with the corresponding compute time shown in the top-right corner of each panel and the extracted charge state overlaid. Panel~\ref{fig:double_dot}(a) shows the steady-state solution obtained using the original \texttt{QArray} method, which relies on numerically solving for the stationary occupation probabilities at each point in gate space. In contrast, panel~\ref{fig:double_dot}(b) displays the charge stability diagram generated by the \texttt{QArray+} stochastic solver in the idealized limit of infinite tunnel rates and zero lead temperature, where the system deterministically relaxes to the minimum-energy charge configuration. Panels~\ref{fig:double_dot}(c)--(g) demonstrate latched charge stability diagrams under finite tunnel rates. In this regime, we assume that each horizontal sweep (each ``row'' of the voltage grid) begins after sufficient relaxation time such that the system returns to its ground-state charge configuration. This assumption enables efficient parallel computation across rows. Panel~\ref{fig:double_dot}(h) shows the fully isolated limit, in which the dots are disconnected from their reservoirs. Here, the user specifies an initial charge state at the lower-left corner, and the state is stochastically propagated across the grid. As no relaxation occurs, each row depends on the evolution of the previous one, and parallelization is therefore not possible. All simulations in Fig.~\ref{fig:double_dot} use $N_r = 1$ and simulate the effect of a charge sensor incorporating $1/f$ noise.

\begin{figure}
    \centering
    \includegraphics{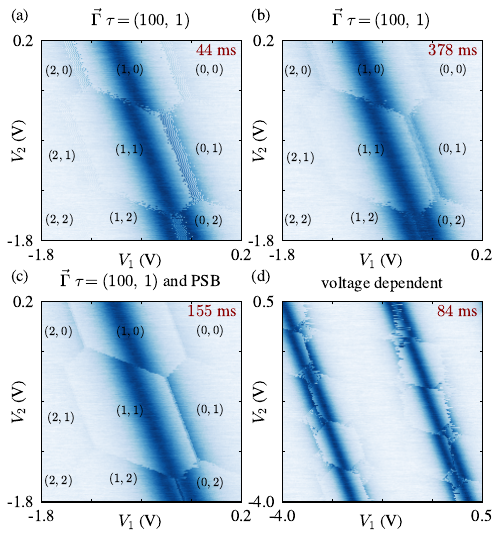}
    \caption{
\textbf{Hole-based charge stability diagrams illustrating how increasing levels of experimental realism---including charge-sensor readout, Pauli spin blockade (PSB), and finite-temperature effects---modify the observed features.} All simulations are performed on a $200 \times 200$ voltage grid.
(a--b) Double-dot charge stability diagrams simulated at $k_B T = 0.04$ for $N_r = 1$ and $N_r = 10$, respectively. Fast tunneling between the first dot and the leads, together with finite-temperature effects near the lead transitions, make loading and unloading comparably likely and lead to stochastic charge-state switching. For $N_r = 1$, this leads to unphysical artifacts in the CSD, while for $N_r = 10$ the fluctuations are captured correctly, resulting in the experimentally observed doubling of the transition.
(c) Double-dot charge stability diagram simulated at $k_B T = 0.01$, demonstrating the effect of Pauli spin blockade, which suppresses the $(1,1) \rightarrow (0,2)$ transition.
(d) Double-dot charge stability diagram incorporating gate-voltage-dependent interdot capacitive coupling and lead tunnel rates.
}    \label{fig:realism}
\end{figure}

To explore the impact of increasing experimental realism, Fig.~\ref{fig:realism} presents simulated charge stability diagrams incorporating Pauli spin blockade (PSB) and finite-temperature effects. 

Panels~\ref{fig:realism}(a--b) display a double-dot diagram computed at a finite temperature of $k_{\mathrm{B}}T = 0.04$, expressed in units of the average dot charging energy. In this simulation, the first dot is strongly coupled to the lead ($\Gamma_1 \tau = 100$), while the second dot is more weakly coupled ($\Gamma_2 \tau = 1$). Near the charge transition, the high tunneling rate to the lead causes the charge state to fluctuate rapidly between $(1, 1)$ and $(0, 1)$. While these individual stochastic events are resolved in the raw simulation, they are typically not observed experimentally due to the finite bandwidth of the measurement setup; instead, the measured signal represents a time average over multiple transitions.

In Panel~\ref{fig:realism}(b), we simulate this experimental limit by increasing the recursion depth $N_r$ to allow for multiple stochastic jumps within a single timestep. The resulting time-averaged signal yields a characteristic ``double-step'' feature frequently observed in experimental measurements (see Appendix~\ref{sec:experimental_csd}). 

Panel~\ref{fig:realism}(c) illustrates the effect of Pauli spin blockade, which suppresses the $(1,1) \rightarrow (0,2)$ transition and manifests as latching at the interdot transition boundary. Finally, Panel~\ref{fig:realism}(d) demonstrates the model's capability to incorporate gate-voltage dependence within the capacitance matrices and tunnel rates, further aligning the simulation with the non-linearities of physical device behavior.

\subsection{Spinless Hubbard model}

The stochastic capacitance model captures non-equilibrium latching, but by construction treats each
integer charge configuration as a classical state and therefore cannot reproduce coherent interdot
hybridization. We next quantify the impact of coherent interdot tunneling in the \emph{equilibrium}
limit using the spinless Hubbard model described above.

\begin{figure}
    \centering
    \includegraphics{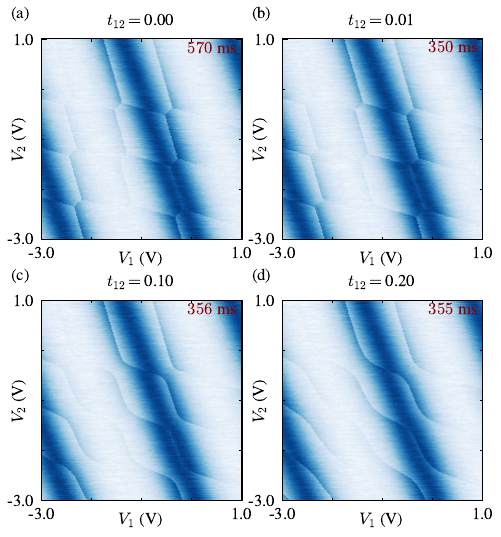}
    \caption{
 \textbf{Tunnel-coupled ground-state (spinless Hubbard) charge stability diagrams for increasing
    interdot tunnel coupling $t_{12}$.}
    Coherent tunneling hybridizes near-degenerate interdot charge configurations, producing avoided
    crossings and fractional charge expectation values that broaden the interdot transition lines as
    $t_{12}$ increases. Each panel shows a $200\times 200$ gate-voltage scan; compute times on CPU
    are indicated in the top-right corners. The larger compute time for the $t_{12}=0$ panel
    reflects one-time just-in-time compilation included in its timing; the remaining panels reuse
    the compiled kernel.}\label{fig:tunnel}
\end{figure}

Figure~\ref{fig:tunnel} shows hole-based double-dot charge-stability diagrams computed at zero
temperature for increasing interdot tunnel coupling $t_{12}$. In the limit $t_{12}=0$
[Fig.~\ref{fig:tunnel}(a)], the result reduces to the classical ground-state picture: the diagram is
piecewise constant with sharp boundaries between integer charge states, yielding the familiar
honeycomb pattern. For finite $t_{12}$ [Fig.~\ref{fig:tunnel}(b--d)], the interdot charge
degeneracies are hybridized. Concretely, near a degeneracy between two configurations connected by a
single interdot hop, e.g.\ $\ket{n_1,n_2}$ and $\ket{n_1-1,n_2+1}$, the effective two-level sector
exhibits an avoided crossing with gap $\sim 2t_{12}$, and the ground-state occupations become
\emph{fractional}: $\langle n_1\rangle$ and $\langle n_2\rangle$ interpolate continuously across the
interdot transition instead of switching discontinuously \cite{Hanson_review}. In the simulated sensor response this
appears as a progressive broadening and rounding of the interdot transition lines and a smoothing of
features around the triple-point regions as $t_{12}$ increases.

An important structural point is that the coherent tunnel Hamiltonian $\hat{H}_t$ conserves total
charge. As a result, only charge configurations within a fixed total-occupation manifold hybridize,
so lead-transition boundaries (which separate different total-charge sectors) remain sharp in this
equilibrium solver. This cleanly isolates the physics of hybridization, producing avoided crossings and charge
delocalization free of any competing dissipative effects. It also motivates the open-system
treatment below, which becomes necessary when finite-rate exchange with reservoirs and
scan-history effects are essential.

\subsection{Open quantum system}
Finally, we combine coherent hybridization with finite-rate dissipative tunneling using the
open-quantum-system model. In this setting, the relevant competition is between (i) the coherent
interdot hybridization scale set by $t_{12}$, (ii) the dissipative lead tunneling rates
$\Gamma_{\mathrm{lead},i}$, and (iii) the effective dwell time per pixel $\tau_{\mathrm{int}}$.
When the characteristic relaxation rate $\Gamma_{\mathrm{relax}}$ set by these dissipative
couplings is fast on the scale of a pixel ($\Gamma_{\mathrm{relax}}\tau_{\mathrm{int}}\gg 1$),
the diagram approaches the equilibrium (Hubbard) limit; when relaxation is slow
($\Gamma_{\mathrm{relax}}\tau_{\mathrm{int}}\lesssim 1$), scan-history dependence emerges and latching-like
metastability becomes visible even in the presence of coherent coupling.

\begin{figure}
    \centering
    \includegraphics{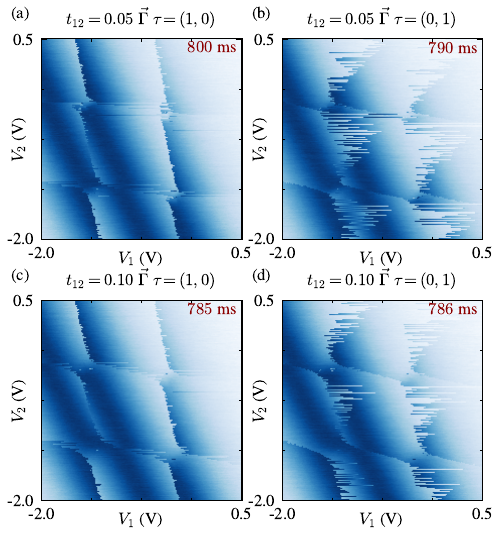}
    \caption{
\textbf{Open-system (Lindblad) charge stability diagrams combining coherent hybridization and
    finite-rate tunneling to leads.}
    The panels compare two interdot tunnel couplings ($t_{12}=0.05$ and $t_{12}=0.10$) and two
    asymmetric reservoir couplings, $\Gamma_{\mathrm{lead}}=(1,0)$ and $\Gamma_{\mathrm{lead}}=(0,1)$,
    illustrating the emergence and suppression of scan-history-dependent latching as the balance
    between coherent tunneling, dissipative tunneling, and pixel dwell time is varied. Each panel
    shows a $200\times 200$ scan; compute times on CPU are indicated.}
    \label{fig:lindblad}
\end{figure}

Figure~\ref{fig:lindblad} demonstrates this crossover using the Lindblad solver in a regime where the
device is \emph{asymmetrically} coupled to the reservoirs. We set the temperature to
$k_{\mathrm{B}}T=0.01$ (in units of the mean charging energy) and evolve the system for a fixed
interval $\tau_{\mathrm{int}}$ per pixel, propagating the quantum state along the scan direction in
parallel mode, i.e., with each row initialized in the instantaneous ground state. To isolate the
interplay between coherent hybridization and reservoir exchange, phonon-mediated incoherent
interdot transitions are disabled in this example ($\Gamma_{\mathrm{phon}} = 0$), so redistribution
between dots occurs only through the coherent tunnel coupling $t_{12}$. With the phonon bath
switched off, the hierarchy of \autoref{eq:timescale_hierarchy} reduces to the single requirement
$\Delta_0 \gg \Gamma_{\mathrm{lead}}$, which is satisfied here: the eigenbasis construction only
requires the coherent scale to dominate all active dissipative rates.

Panels~\ref{fig:lindblad}(a,c) show the case where only dot~1 is coupled to a reservoir,
$\Gamma_{\mathrm{lead}}=(1,0)$, while panels~\ref{fig:lindblad}(b,d) couple only dot~2,
$\Gamma_{\mathrm{lead}}=(0,1)$. When the lead is attached to dot~1, the system can efficiently load
and unload charge through that dot; combined with coherent hybridization, this provides an effective
relaxation pathway for the two-dot manifold and yields comparatively smooth, near-equilibrated
transition features. In contrast, when the lead is attached to dot~2, charge exchange occurs through
a different physical bottleneck relative to the scan axes, and the system can become trapped in metastable charge configurations over extended
scan ranges, producing jagged, scan-direction-dependent transition features characteristic of
latching. Increasing the coherent coupling from $t_{12}=0.05$ to $t_{12}=0.10$
[Fig.~\ref{fig:lindblad}(b) vs.\ (d)] suppresses these non-equilibrium artifacts by increasing the
hybridization-induced connectivity between relevant charge configurations, thereby accelerating
relaxation through the lead-coupled dot and pushing the diagram toward the equilibrium limit.

Because the Lindblad solver propagates the quantum state across the scan and
samples stochastic quantum jumps, it is computationally more expensive than the equilibrium
ground-state solver for the same grid resolution; nevertheless, the JAX implementation keeps these
simulations tractable and enables accelerator execution for large-scale dataset generation.

\section{Benchmarking} \label{sec:benchmarking}

\begin{figure*}
    \centering    \includegraphics[width = \textwidth]{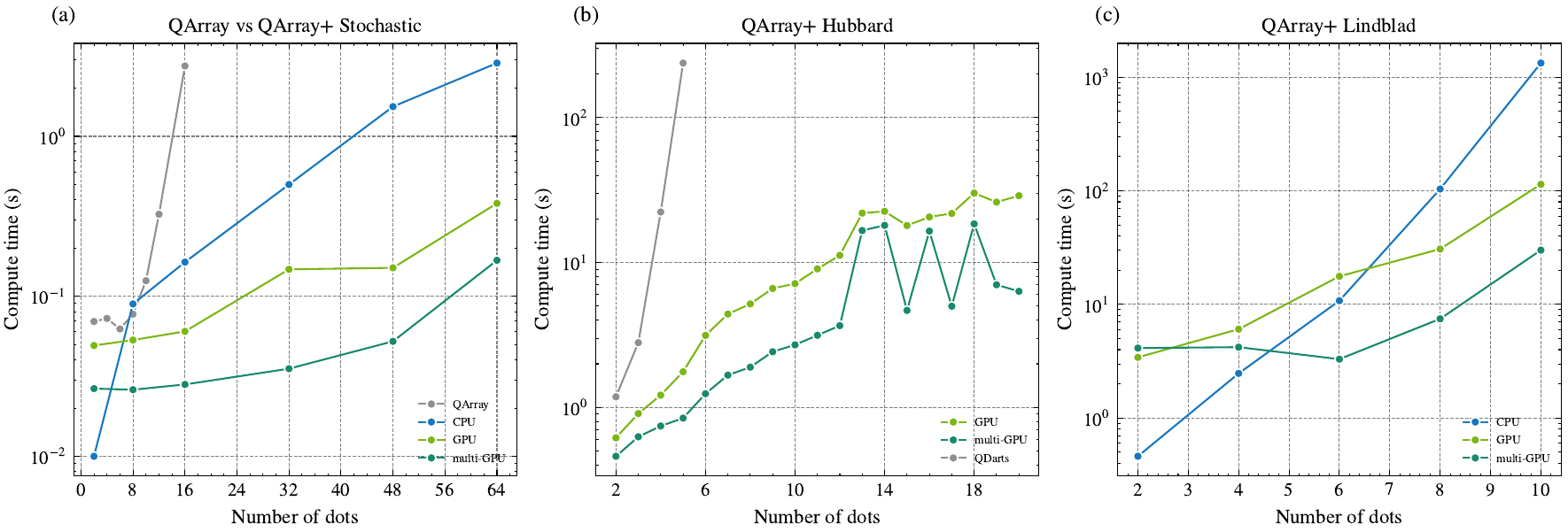}
 \caption{\textbf{Benchmarks for QArray+ on latching, tunnel coupling and Lindblad features.} All benchmarks use a 100 × 100 charge-stability-diagram resolution on an 8 × NVIDIA H100 (80 GB HBM3) node. (a) QArray baseline versus QArray+ latching on CPU, single GPU, and 8-GPU, showing that QArray+ on GPU and multi-GPU outperforms the baseline at all dot counts scaling up to 64 dots; multi-GPU maintains sub-second runtimes throughout. (b) Tunnel-coupling benchmark comparing QArray+ (single GPU and multi-GPU) against QDarts, which exceeds 200s at 5 dots while QArray+ remains tractable to 20 dots. (c) Lindbladian benchmark; multi-GPU provides an order-of-magnitude speedup over CPU at 10 dots.
}

\label{fig:scaling}
\end{figure*}

All the CSDs displayed so far were computed using the CPU of an M3 MacBook Pro; the compute time on this hardware is superimposed in the top right-hand corner. We now benchmark the scaling
of \texttt{QArray+} to larger arrays and quantify accelerator speedups on a server-class system.

All benchmarks were run on a node equipped with 8$\times$ NVIDIA H100 SXM5 80\,GB GPUs and
Intel Xeon Platinum 8480C CPUs. Unless stated otherwise, each point
corresponds to the profiler time required to compute a full $100\times 100$ charge-stability
diagram (CSD) after JIT compilation and warm-up. Since each benchmark
computes a fixed-resolution scan, the reported times scale approximately linearly with the number of
pixels, allowing extrapolation to other grid sizes (e.g., $200\times 200$ is $\approx 4\times$ more work than $100 \times 100$) as seen in the Appendix.

Figure~\ref{fig:scaling} summarizes three representative workloads corresponding to the main
simulation heads used in \texttt{QArray+}. We highlight the following subgraphs:

\paragraph{Stochastic latching (capacitance Markov-jump solver).}
Fig.~\ref{fig:scaling}(a) compares the original \texttt{QArray} baseline implementation to the \texttt{QArray+} latching solver on CPU, single GPU, and multi-GPU execution.
The baseline shows a rapid growth in runtime with dot count and becomes impractical beyond
$\sim$16 dots over the tested range, consistent with the exponential growth of the underlying
state space in steady-state formulations. In contrast, the \texttt{QArray+} latching solver exhibits
the expected polynomial scaling in $n_{\mathrm{dot}}$ and remains tractable out to 64 dots.
At the largest system size shown, the CPU backend requires $\sim$2.8\,s per $100\times 100$ scan,
while a single H100 reduces this to $\sim$0.4\,s and 8$\times$H100 reduces it further to
$\sim$0.17\,s, demonstrating substantial end-to-end speedups from accelerator execution. 

\paragraph{Tunnel-coupled ground state (spinless Hubbard head).}
Fig.~\ref{fig:scaling}(b) benchmarks the tunnel-coupled ground-state solver against \texttt{QDarts}.
While \texttt{QDarts} becomes expensive already for $\gtrsim 5$ dots (hundreds of seconds per scan in
this configuration) and extrapolates to prohibitive runtimes at larger $n_{\mathrm{dot}}$,
\texttt{QArray+} remains in the few-second to few-tens-of-seconds regime up to 20 dots.
This improvement is enabled by the fixed-size truncated charge basis and the use of sparse linear
algebra (Lanczos-type eigensolvers), which avoid explicit enumeration and diagonalization in the
full exponentially large charge basis. Across the tested range, sparse multi-GPU execution provides
an additional speedup over a single GPU, with the benefit depending on the effective sparsity of the
tunnel graph and the resulting per-point Hamiltonian construction and eigensolve cost.

\paragraph{Open-system Markov jumps (Lindblad MCWF eigenjump head).}
Fig.~\ref{fig:scaling}(c) reports performance of the open-system Lindblad solver. The accelerator execution
substantially extends the accessible regime of this fully physics-informed solver: at 10 dots, the CPU runtime reaches $\sim$1.3$\times 10^3$\,s
($\sim$22\,min) per scan, whereas a single H100 reduces this to $\sim$110\,s
($\sim$2\,min), and 8$\times$H100 reduces it to $\sim$30\,s.

Overall, these benchmarks show that \texttt{QArray+} supports high-throughput dataset generation
across multiple physical regimes, and that JAX-based accelerator execution, especially multi-GPU
sharding, yields order-of-magnitude reductions in end-to-end runtime for the workloads most relevant
to large-scale simulation of quantum-dot arrays.




\section{Conclusion}
We have introduced \texttt{QArray+}, a versatile simulator for modeling non-equilibrium charge dynamics in tunnel-coupled semiconductor quantum-dot arrays. By providing a quantum open-system description based on Lindblad master-equation evolution, the framework enables a unified treatment of coherent dynamics and dissipative tunneling to leads. This approach allows for the generation of charge-stability diagrams (CSDs) that simultaneously capture complex phenomena such as latching and interdot tunnel coupling. 

In regimes where coherent effects are less dominant, \texttt{QArray+} allows for a transition to a stochastic model characterized by polynomial rather than exponential scaling. This efficiency enables the simulation of large-scale quantum-dot arrays that would be computationally prohibitive using conventional methods. Consequently, we anticipate that this framework will serve as a robust ``digital twin'' for the offline testing and verification of large-scale devices.

The framework naturally incorporates gate-voltage-dependent capacitances, Pauli spin blockade, finite-temperature effects, realistic sensor readout and noise models. These features make \texttt{QArray+} suited for integration into automated tuning pipelines and the generation of large-scale synthetic datasets for machine-learning applications. As quantum-dot devices grow in complexity and fast measurements become widespread, the ability to bridge tunnel-coupling-induced hybridization with dissipative non-equilibrium dynamics will be essential. The methods presented here lay the groundwork for scalable, realistic simulation tools designed to support the next generation of semiconductor quantum processors. Future work using \texttt{QArray+} will also enable rapid data generation for training autonomous tuning pipelines, which are crucial for realizing scalable semiconductor quantum devices.

\section*{Acknowledgments}

P.V. performed this research as part of an internship at NVIDIA.
N.A. acknowledges support from the European Research Council (grant agreement 948932), the Royal Society (URF-R1-191150) and the United States Army Research Office under Award No. W911NF-24-2-0043. 
This research was also supported by the European Union through the Horizon 2020 research and innovation program under the Grant Agreement No. 951852 (QLSI) and the Horizon Europe Framework Program under grant agreement No. 101069515 (IGNITE).
Views and opinions expressed are, however, those of the authors only and do not necessarily reflect those of the European Union, Research Executive Agency or UK Research \& Innovation. Neither the European Union nor UK Research \& Innovation can be held responsible for them. P.V. is supported by the United States Army Research Office under Award No. W911NF-21-S-0009-2.

A.P., B.v.S., and M.V. acknowledge support from the NWO through a Vici grant (VI.C.242.031) and the National Growth Fund program Quantum Delta NL (grant NGF.1582.22.001). This research was sponsored in part by the Army Research Office (ARO) under Award No. W911NF-23-1-0110 and by The Netherlands Ministry of Defense under Award No.~QuBits R23/009. The views, conclusions, and recommendations contained in this document are those of the authors and are not necessarily endorsed nor should they be interpreted as representing the official policies, either expressed or implied, of the Army Research Office (ARO) or the U.S. Government, or The Netherlands Ministry of Defense. The U.S. Government and The Netherlands Ministry of Defense are authorized to reproduce and distribute reprints for Government purposes notwithstanding any copyright notation herein.

B.v.S. thanks Francesco Borsoi for fruitful discussions about latching.

\section*{Code availability}
The \texttt{QArray+} Python package will be available on GitHub.

\section*{Author contributions}

A.P. and B.v.S. developed the stochastic capacitance model. P.V. developed and implemented the spinless Hubbard model and the tunnel coupling in collaboration with B.v.S. E.D.N. and R.M. developed the charge truncation, supervised by P.V. and B.v.S. P.V. developed the open quantum system code under the supervision of T.P., B.K., N.A., and B.v.S. P.V. developed the multi-GPU acceleration under the supervision of T.P., B.K., and N.A. M.V. supervised A.P. and B.v.S.
The manuscript was written by P.V. and B.v.S. with helpful contributions and comments from all authors.
\section*{Declarations}

M.V. is a founding advisor of Groove Quantum BV and declares equity interests. N.A. declares an interest as a founder
of QuantrolOx, which develops machine learning-based
software for quantum control. All remaining authors declare no conflicts of interest.

\FloatBarrier
\bibliographystyle{unsrt} 
\bibliography{references}
\appendix
\newpage

\section{Computation complexity}\label{app:complexity}
Here we discuss the complexity of computing an $N \times N$ charge stability diagram using the stochastic capacitance model for a system consisting of $n_{\mathrm{dot}}$ dots. Per sub-interval, there are $2n_{\mathrm{dot}}$ loading/unloading transitions and $n_{\mathrm{dot}}(n_{\mathrm{dot}} - 1)$ interdot transitions, i.e., $n_{\mathrm{dot}}(n_{\mathrm{dot}} + 1)$ transitions per sub-interval and hence $N_r\, n_{\mathrm{dot}}(n_{\mathrm{dot}} + 1)$ candidate charge transitions per pixel. For each of these $\mathcal{O}(N_r\, n_{\mathrm{dot}}^2)$ transitions it is necessary to evaluate the electrostatic energy according to \autoref{eq:U}, which requires multiplying an $n_{\mathrm{dot}} \times n_{\mathrm{dot}}$ matrix by a vector of length $n_{\mathrm{dot}}$, an operation with complexity $\mathcal{O}\!\left(n_{\mathrm{dot}}^{2}\right)$. The per-pixel cost is therefore $\mathcal{O}\!\left(N_r\, n_{\mathrm{dot}}^{4}\right)$, and the overall computational complexity of the charge stability diagram is $\mathcal{O}\!\left(N^2 N_r\, n_{\mathrm{dot}}^{4}\right)$. The computational complexity of this algorithm therefore scales more favorably with the number of dots than that of steady-state methods. We note that caching $\mathbf{C}_{dd}^{-1}\vec{Q}$ at each pixel reduces the cost of evaluating the energy change of a single transition to $\mathcal{O}(n_{\mathrm{dot}})$, which would tighten the per-pixel scaling to $\mathcal{O}\!\left(N_r\, n_{\mathrm{dot}}^{3}\right)$.

This polynomial scaling represents a dramatic improvement over steady-state solvers, which typically must evaluate the electrostatic energy of at least $2^{\,n_{\mathrm{dot}}}$ distinct charge states. Therefore, their complexity scales as at least $\mathcal{O}\!\left(2^{\,n_{\mathrm{dot}}}\right)$.

\begin{figure}
    \centering
    \includegraphics[width=\columnwidth]{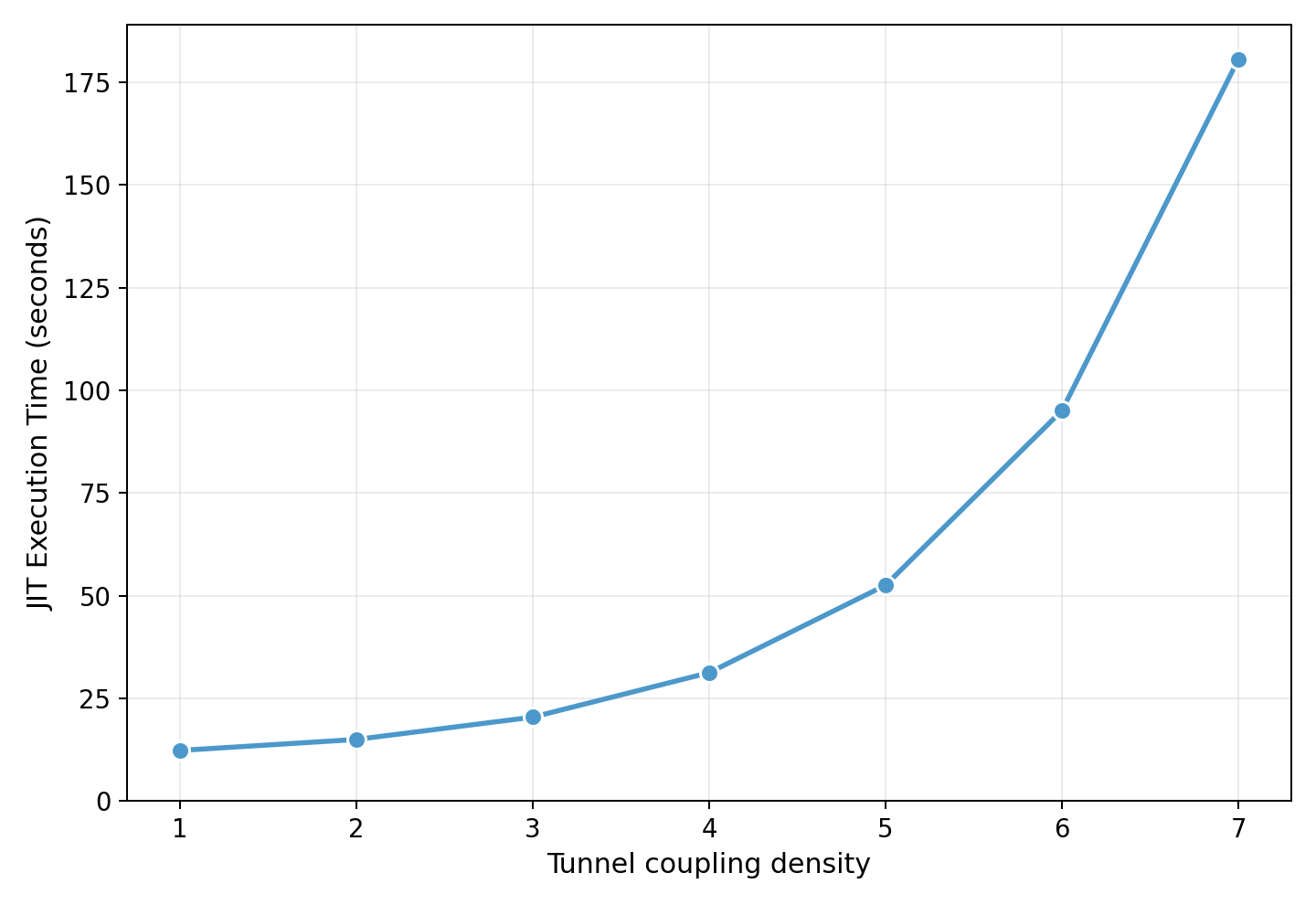}
    \caption{\textbf{JIT execution time vs.\ tunnel-coupling density for a 20-dot array (64$\times$64 grid, multi-GPU).}
    The JAX just-in-time (JIT) execution time increases as the tunnel-coupling density grows, consistent with a larger effective connectivity that increases the number of tunnel-coupled transitions (and, consequently, the amount of work per voltage point) even when distributing the computation across multiple GPUs.}
    \label{fig:appendix-jit-vs-density}
\end{figure}

As shown in Fig.~\ref{fig:appendix-jit-vs-density}, JIT execution time rises rapidly at high tunnel-coupling density. In practice, dense tunnel graphs increase the number of nonzero couplings $t_{ij}$ (and associated rates) that must be evaluated at each voltage point, which can dominate runtime even under multi-GPU parallelism. This benchmark highlights the importance of (i) exploiting JAX/XLA fusion and (ii) using multi-GPU/multi-node execution for high-throughput dataset generation in regimes with strong interdot connectivity.

\section{Comparison of simulators with QArray latching} \label{app:comparison}

As per \autoref{tab:comparison}, the original \texttt{QArray} \cite{van_Straaten_2024} and QDFlow \cite{buterakos2025qdflowpythonpackagephysics} have the ability to simulate latching in an unphysical fashion. They neglect the gate-voltage dependence of the relaxation rate and the possibility of relaxation through intermediate charge states. 

\begin{figure}[h]
	\centering
	\includegraphics{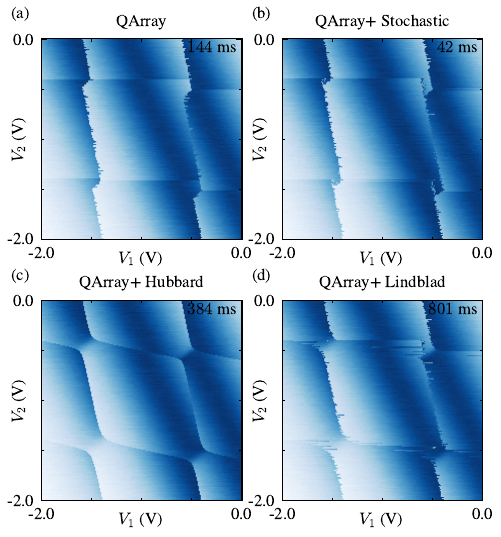}
    \caption{\textbf{Double-dot charge stability diagrams for reservoir couplings $\vec{\Gamma}\tau = (1, 0)$, computed using four different methods.} (a)~The original \texttt{QArray} latching implementation, and (b)--(d) the \texttt{QArray+} stochastic, spinless Hubbard, and Lindblad (open-system) models, respectively, for the same double-dot system. Dot 1 exchanges charge rapidly with the reservoir while dot 2 is isolated. Compute times for each $200 \times 200$ scan are indicated in the top-right corners.}
    \label{fig:comparison_good}
\end{figure}

In \autoref{fig:comparison_good}, we simulate the same double quantum dot (DQD) 
system as in \autoref{fig:models}, though here the original \texttt{QArray} 
is used to simulate charge latching. The system is configured with tunneling
rates $\vec{\Gamma} \tau = (1, 0)$, such that dot 1 readily exchanges
charges with the reservoir while dot 2 remains isolated. Consider the latching that occurs when the $(1, 1) \to (1, 0)$ transition 
is hindered by a negligible loading/unloading rate. In the standard 
\texttt{QArray} simulation, the system remains trapped in the $(1, 1)$ state 
until it crosses the $(1, 0) \to (0, 0)$ unloading boundary. In contrast, 
the more physically grounded \texttt{QArray+} simulation remains in $(1, 1)$ 
only until reaching the \textit{extended} $(1, 1) \to (0, 1)$ unloading line. 
At this point, a rapid two-step transition $(1, 1) \to (0, 1) \to (1, 0)$ 
becomes possible, facilitated by the fast loading/unloading of Dot 1 and 
the underlying inter-dot transition rates.

\begin{figure}[h]
	\centering
	\includegraphics{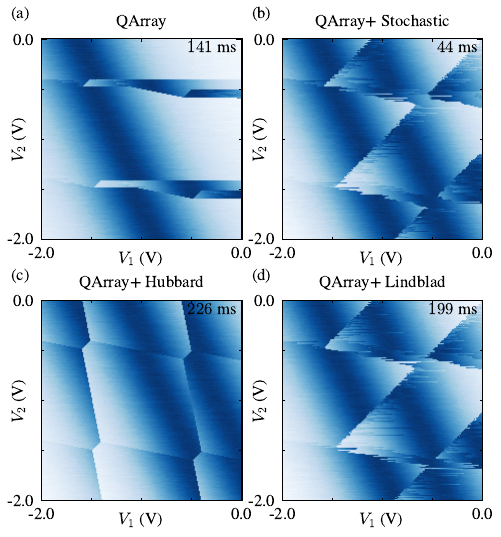}
    \caption{\textbf{Double-dot charge stability diagrams for reservoir couplings $\vec{\Gamma}\tau = (0, 0.1)$, computed using four different methods.} (a)~The original \texttt{QArray} latching implementation, and (b)--(d) the \texttt{QArray+} stochastic, spinless Hubbard, and Lindblad (open-system) models, respectively. Here dot 1 is fully decoupled from its reservoir and dot 2 is weakly coupled. Compute times for each $200 \times 200$ scan are indicated in the top-right corners.}
    \label{fig:comparison_bad}
\end{figure}

In \autoref{fig:comparison_bad} we simulate the same double dot system, but now with $\vec{\Gamma} \tau = (0, 0.1)$. With these rates, the \texttt{QArray} latching simulation breaks down entirely: in \autoref{fig:comparison_bad}(a), the charge state remains latched within narrow horizontal bands that streak across the entire sweep, bearing little resemblance to the extended latched regions produced by the physically grounded models in panels (b) and (d). In the \texttt{QArray+} simulation, by contrast, the vanishing rate of dot 1 prevents the $(2, 1) \to (1, 1)$ transition from occurring, so instead there is an \textit{extended} interdot transition corresponding to $(2, 1) \to (1, 2) \to (1, 1)$.

\section{Discrete-Time Steady-State Occupation}
\label{app:steady_state}

In this appendix, we derive the discrete-time steady-state occupation probability utilized in the latching model. We consider a single quantum dot coupled to one electronic reservoir and restrict our attention to two charge states, $\vec{n}_0$ and $\vec{n}_1$. The dimensionless electrostatic energy change associated with a loading event is defined as
\begin{equation}
    x \equiv \beta \Delta U, \qquad \beta = (k_B T)^{-1}.
\end{equation}

\subsection{Transition Rates and Probabilities}
Electron exchange with the reservoir is governed by Fermi-Dirac statistics. The loading ($\Gamma_{0\to1}$) and unloading ($\Gamma_{1\to0}$) rates are given by
\begin{subequations}
\begin{align}
    \Gamma_{0\to1} &= \Gamma f(x), \\
    \Gamma_{1\to0} &= \Gamma f(-x),
\end{align}
\end{subequations}
where $f(x) = (1 + e^{x})^{-1}$ is the Fermi function and $\Gamma$ denotes the bare tunnel rate. 

To model the system over discrete integration time steps of duration $\tau$, we define the probabilities $A$ and $B$, which represent the probability of the system remaining in charge state 0 and 1, respectively, during the interval $\tau$:
\begin{subequations}
\begin{align}
    A &= \exp\left[-\Gamma \tau f(x)\right], \\
    B &= \exp\left[-\Gamma \tau f(-x)\right].
\end{align}
\end{subequations}

\subsection{Steady-State Derivation}
The discrete-time evolution over one integration step is described by the stochastic transition matrix $\mathbf{T}$:
\begin{equation}
    \mathbf{T} = \begin{pmatrix} A & 1-B \\ 1-A & B \end{pmatrix}.
\end{equation}

The stationary occupation probability $\pi_1$ satisfies the eigenvalue condition $\boldsymbol{\pi} = \mathbf{T}\boldsymbol{\pi}$, where $\boldsymbol{\pi} = (1-\pi_1, \pi_1)^T$. This leads to the balance equation:
\begin{equation}
    \pi_1 = (1-A)(1-\pi_1) + B \pi_1.
\end{equation}
Solving for $\pi_1$ yields the steady-state occupation:
\begin{equation}
    \pi_1(x) = \frac{1 - A}{2 - A - B}.
\end{equation}
The behavior of the system is fundamentally governed by the dimensionless ``bandwidth'' parameter $\Gamma \tau$, which determines the deviation from the continuous-time limit.
\begin{figure}
    \centering
    \includegraphics{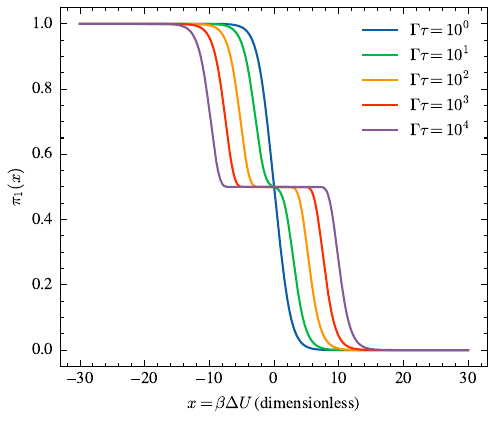}
    \caption{\textbf{Steady-state occupation $\pi_1$ as a function of the dimensionless electrostatic energy $x$.} The curves illustrate how the transition broadens and flattens with the parameter $\Gamma \tau$.}
    \label{fig:steady_state_plot}
\end{figure}

\section{Experimentally measured CSD} \label{sec:experimental_csd}

\begin{figure}
	\centering
    \includegraphics{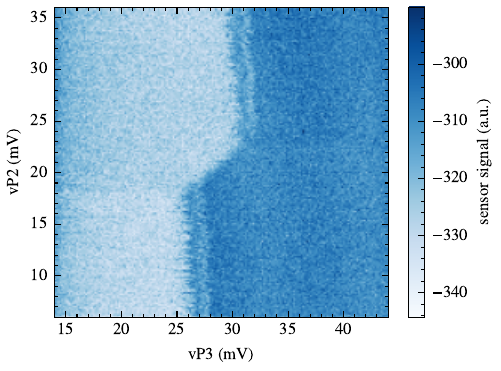}
    \caption{\textbf{Experimentally measured CSD showing a double transition line, adapted from Ref.~\cite{van_Riggelen_Doelman_2024}.}}
    \label{fig:experimental_csd}
\end{figure}

\autoref{fig:experimental_csd} shows an experimentally measured charge stability diagram
exhibiting a double transition line at a dot--reservoir charge transition. This is the same
feature reproduced by the stochastic capacitance model at finite temperature and fast lead
tunneling in \autoref{fig:realism}(b), where resolving multiple stochastic transitions per pixel
($N_r = 10$) yields the time-averaged ``double-step'' signal discussed in \autoref{sec:results}.

\end{document}